\documentclass[review,a4paper,12pt, number]{elsarticle}
\usepackage{amssymb, mathtools}
\usepackage{algpseudocode,algorithm}
\usepackage{chemarrow}
\usepackage[margin=1.8cm]{geometry}
\usepackage{color}
\graphicspath{{plots/}}

\newcommand{\wrt}[1]{\mathrm{d}{#1}}
\newcommand{\bs}[1]{\boldsymbol{#1}}

\usepackage{etoolbox}
\makeatletter
\patchcmd{\ps@pprintTitle}{\footnotesize\itshape
       Preprint submitted to \ifx\@journal\@empty Elsevier
       \else\@journal\fi\hfill\today}{\relax}{}{}
\makeatother

\begin{document}

\begin{frontmatter}

\title{Robustly simulating biochemical reaction kinetics using multi-level Monte Carlo approaches}
\author{Christopher Lester$^\dagger$, Christian A. Yates$^*$, Ruth E. Baker$^\dagger$}

\address{$^\dagger$Mathematical Institute, Woodstock Road, Oxford, OX2 6GG, UK and $^*$Department for Mathematical Sciences, Claverton Down, Bath, BA2 7AY, UK.}

\begin{abstract}
In this work, we consider the problem of estimating summary statistics to characterise biochemical reaction networks of interest. Such networks are often described using the framework of the Chemical Master Equation (CME). For physically-realistic models, the CME is widely considered to be analytically intractable. A variety of Monte Carlo algorithms have therefore been developed to explore the dynamics of such networks empirically. Amongst them is the multi-level method, which uses estimates from multiple ensembles of sample paths of different accuracies to estimate a summary statistic of interest. {In this work, we develop the multi-level method in two directions: (1) to increase the robustness, reliability and performance of the multi-level method, we implement an improved variance reduction method for generating the sample paths of each ensemble; and (2) to improve computational performance, we demonstrate the successful use of a different mechanism for choosing which ensembles should be included in the multi-level algorithm.}
\end{abstract}


\end{frontmatter}

\section{Introduction} \label{__label__134bf72532374159b9f5b8312237546a}
Stochastic approaches are commonly used to model a wide variety of biological processes~\citep{__ref__c9c0acbfcc1e466796f203c599f87a48,__ref__8a7bc56c9dfd469dbc93e2e8bbf7172b,__ref__fa6e3de85e0b4412ba3096011a9acb70}. This work is concerned with individual-based models that represent the behaviour of single particles or organisms within a biological system of interest. We use discrete reaction events to describe the interactions between individual particles; the time evolution of the system is described by the Chemical Master Equation (CME)~\citep{__ref__37a69023cd02435cb5c677546fc5d00e}. Except for a small number of special cases, the CME is both analytically and numerically intractable~\citep{__ref__3e0b45aec9df4e05b2f3e8165dab4878}, and Monte Carlo simulation is often used to explore the model dynamics instead. Typically, a number of sample paths are generated, and used to estimate relevant summary statistics.

The multi-level method was described for stochastic differential equations by \citet{__ref__dbd754638429454ab66f23e25ec8f0be}, and in 2012, \citet{__ref__5f82fa75a5314b0484ebeb52de9e2750} developed an approach suitable for biochemical reaction networks. The multi-level method produces point estimators in a cost-effective manner by combining a hierarchy of tau-leap estimators of differing accuracy in a telescoping sum. When compared with traditional simulation methods, the multi-level method has been shown to be capable of reducing the CPU time taken to estimate summary statistics by orders of magnitude~\citep{__ref__5f82fa75a5314b0484ebeb52de9e2750}. A wide body of subsequent research has focussed on generalising the multi-level method used by Anderson and Higham~\citep{__ref__cefabb61e12c4615ba29a64649c34236, __ref__25cacfb38a074fdf9ed2a1329b16f350, __ref__32e9156a6baf4b31a6cfc19f7b417e76}. A number of general performance improvements~\citep{__ref__b7cd69d78ffc4c2ba54f2f3e26b5d981} have also been described, but current, state-of-the-art implementations of the multi-level method still rely heavily on user input and configuration. Additionally, a number of unusual implementation issues remain unresolved~\citep{__ref__25cacfb38a074fdf9ed2a1329b16f350, __ref__b7cd69d78ffc4c2ba54f2f3e26b5d981}. 

This work has two aims. Firstly, to improve the efficiency and reliability of the multi-level method, we change the procedure by which the terms in the aforementioned telescoping sum are computed. Secondly, we provide further improvements in computational performance by including R-leap estimators, instead of tau-leap estimators, in the telescoping sum.

This work is arranged as follows: in Section \ref{__label__e32e9d19f65f477a83d3ba66faafa7cf} we review background material, so that the multi-level method can be described in Section \ref{__label__471d7eaf487749a0885bc893309e52bd}. In Sections \ref{__label__788405dd7a434bb5acdbf2a30ef377af} and \ref{__label__576de80bab724c3284173a8363d971e7} we develop a new and reliable implementation of the multi-level method (the first aim of our work). Our new simulation method is compared numerically with a traditional approach in Section \ref{__label__439a9eb968c54826b8730e3976b4d497}. A new multi-level framework is then set out and tested numerically in Section \ref{__label__4f2afd7baa2b4732a8ee837f2fc214bf} (our second aim). Conclusions are then drawn in Section \ref{__label__081a6f79524644a9b93459bfc2788cf2}. 

\section{Stochastic biochemical networks} \label{__label__e32e9d19f65f477a83d3ba66faafa7cf}
We study a biochemical network comprising $N$ species, $S_1, \dots, S_N$, that may interact through $M$ reaction channels, $R_1$,$\dots$,$R_M$. In this work, we will assume that the system of interest is well-stirred, and we describe the dynamics of the biochemical network using the CME~\citep{__ref__3e0b45aec9df4e05b2f3e8165dab4878}. At time $t$, the population, or copy number, of species $S_i$ is denoted by $X_i(t)$, and the state vector, $\boldsymbol{X}(t)$, is given by \begin{equation}
\boldsymbol{X}(t) \coloneqq \left[X_1(t), \dots, X_N(t)\right]^T. \label{__label__31ed770f67774bfbb23fdf706bf29500} \end{equation}

We associate two quantities with each reaction channel $R_j$. The first is the stoichiometric or state-change vector, \begin{equation}
\boldsymbol{\nu}_j \coloneqq \left[\nu_{1j}, \dots, \nu_{Nj}\right]^T,  \label{__label__adb913e196a74e2d9817642b922f0226}
\end{equation}
where $\nu_{ij}$ is the change in the copy number of $S_i$ caused by reaction $R_j$ taking place. The second quantity is the propensity function, $p_j(\boldsymbol{X}(t))$. For infinitesimally small $\wrt{t}$, the rate $p_j(\boldsymbol{X}(t))$ is defined as follows: \begin{equation*}
p_j(\boldsymbol{X}(t))\wrt{t} \coloneqq \mathbb{P}\left[R_j \text{ occurs in } [t, t + \wrt{t}) \right].
\end{equation*} Since the system is well-stirred, the reaction activity can, for example, be modelled with mass action kinetics. Where mass action kinetics are used, the propensity function of reaction $R_j$, $p_j$, is proportional to the number of possible combinations of reactant particles in the system~\citep{__ref__3e0b45aec9df4e05b2f3e8165dab4878}.

\subsection{The random time change representation}
In order to analyse and understand the CME framework mathematically, the random time change representation (RTCR) is now set out. The RTCR is a path-wise representation of the \emph{same} stochastic process as the CME, and was first described by \citet{__ref__93139130af66492c89b2da93df8e8526}. The RTCR describes the dynamics of our chosen biochemical network by using a set of inhomogeneous Poisson processes. We first describe the inhomogeneous Poisson process framework, and then provide the RTCR. 

Consider a (homogeneous) Poisson process of fixed rate $\lambda$, labelled as $\mathcal{Y}^\lambda$. Then further suppose we have a Poisson process of unit rate, $\mathcal{Y}^1$. As Poisson processes count the number of `arrivals' over time, they can be compared by considering the distribution of the number of arrivals by some time $t$. If $\mathcal{Y}^\lambda(0,t)$ and $\mathcal{Y}^1(0,t)$ represent the number of arrivals over the time interval $(0,t]$ in each of our Poisson processes, then there is an equality in distribution, that is $\mathcal{Y}^\lambda(0,t) \sim \mathcal{Y}^1(0, \lambda \cdot t)$. It is therefore possible to re-scale time to transform a unit rate Poisson process to one of arbitrary (but known) rate. We will now allow $\lambda$ to change its value. The value of $\lambda$ can change with time, or, alternatively, depending on the arrivals of the Poisson process at earlier times. The inhomogeneous Poisson process formalism allows the rate parameter, $\lambda$, to change dynamically.

The number of times reaction $R_j$ (for $j = 1, \dots, M$) takes place (`fires') over the time interval $(0, T]$ is given by a inhomogeneous Poisson counting process \begin{equation*}
 \mathcal{Y}_j \left(0, \int_0^T p_j(\boldsymbol{X}(t)) \wrt{t}\right),
\end{equation*} where $\mathcal{Y}_j$ is a unit-rate Poisson process, and $\mathcal{Y}_j(\alpha, \beta)$ is defined as \begin{equation}
\mathcal{Y}_j(\alpha, \beta) \coloneqq \text{\# of arrivals in } (\alpha, \beta].
\end{equation} Every time reaction $R_j$ occurs, the state vector (see Equation \eqref{__label__31ed770f67774bfbb23fdf706bf29500}), is updated by adding the appropriate stoichiometric vector (see Equation \eqref{__label__adb913e196a74e2d9817642b922f0226}) to it. Therefore, by considering all possible reactions over the time interval $(0,T]$, we can determine the state vector at time $T$ as \begin{equation}
 \boldsymbol{X}(T) = \boldsymbol{X}(0) + \sum_{j=1}^{M}  \mathcal{Y}_j \left(0,\int_0^T p_j(\boldsymbol{X}(t)) \wrt{t}\right) \cdot \nu_j. \label{__label__0ec1364809bb422f9cb945e0db5d4701}
\end{equation}

{As highlighted in the introduction, analytical solutions of the CME can only be obtained for a relatively small number of special cases. Approximate solutions to the CME might also be obtainable under certain circumstances. For further information, we direct readers to \citet{__ref__31cfa34968c84ce9be58a56caac7157e}. Monte Carlo simulation is, therefore, commonly used as an alternative to understand the system's features and properties~\citep{__ref__3e0b45aec9df4e05b2f3e8165dab4878}.}

We start by discussing three methods for generating suitable sample paths: the direct, tau-leap and R-leap methods. Typically, these methods are used to generate an ensemble of $\mathcal{N}$ sample paths, and the sample paths are then used to estimate summary statistics that characterise the system.

\subsection{Gillespie Direct Method} \label{__label__6fd124019bec45a8bc6d5c32ae94030a}
A variety of methods can be used to generate the sample paths, but the Gillespie Direct Method (DM) is the most widely-used algorithm~\citep{__ref__3e0b45aec9df4e05b2f3e8165dab4878, __ref__85bbd2b5b7ad4596ad0a5acfb3b44152, __ref__9e06f02ea6d8426bb0cae90cddbbe6e7}. The DM is a serial algorithm in the sense that sample paths are generated by simulating individual reactions in the order in which they take place.  A key benefit of the DM is that it can be implemented in a straightforward manner, as shown in Algorithm \ref{__label__c0331a4be21941daa64b9da1f158abd1}. However, because every reaction that takes place in a sample path is simulated individually, the DM requires a relatively high level of computational resources. {This could render the DM unsuitable for very large reaction networks, or where detailed parameter sweeps or model inference needs to be undertaken.}

\begin{algorithm}[bth]
\caption{The DM. This simulates a single sample path.\protect\vphantom{$A_A^A$}}
\label{__label__c0331a4be21941daa64b9da1f158abd1}
 \begin{algorithmic}[1]
  \Require initial conditions, $\bs{X}(0)$, and terminal time, $T$.\protect\vphantom{$A_A^A$}
  \State set $\bs{X} \leftarrow \bs{X}(0)$ and set $t \leftarrow 0$
  
  \Loop
  \State for each $R_j$, calculate propensity values $p_j$ and set $p_0 \leftarrow \sum_{j=1}^M p_j $
  \State set $\Delta \leftarrow \text{Exp}(p_0)$  
  \If{$t + \Delta > T$}
  \State \textbf{break}
  \EndIf
  \State choose reaction $R_k$ to fire next: $R_k$ fires with probability $p_k / p_0$ 
  \State set $\bs{X} \leftarrow \bs{X} + \bs{\nu}_k$, and set $t \leftarrow t + \Delta$
  \EndLoop
 \end{algorithmic}
\end{algorithm}

The summary statistics estimated using Monte Carlo simulation contain a statistical error. This arises as we have generated only a subset of the  possible sample paths: with a different ensemble of realisations, the estimate, $\widehat{\mathcal{Q}}$, will be slightly different, and therefore estimates have an inherent uncertainty to them. More precisely, if the sample variance of the chosen summary statistic is $\mathcal{V}$, and $\mathcal{N}$ sample values have been used to estimate the summary statistic, then the \emph{estimator variance} is given by $\mathcal{V} / \mathcal{N}$. The estimator variance can be used to construct a confidence interval~\citep{__ref__b7cd69d78ffc4c2ba54f2f3e26b5d981}. 

Approximate simulation methods can be implemented to reduce the computational resources required for Monte Carlo simulation. In short, the dynamics of a chosen biochemical reaction network are simplified so that fewer computational resources are required to simulate each sample path. Examples of such approximation schemes include the widely-used tau-leap~\citep{__ref__c646c0890ccf4a59835f2df5c824c223}, and R-leap~\citep{__ref__d45dab7799794c30abe917c6253324aa} methods. We discuss both the tau- and R-leap methods in turn. 

\subsection{The tau-leap method}
The tau-leap method generates approximate sample paths efficiently by taking time-steps of length $\tau$ through time, and firing multiple reactions during each time-step. We will work with a fixed choice of time-step, $\tau$. During each time-step, the key assumption we make is that the reaction propensities are constant. It can then be shown that the number of times that reaction channel $R_j$ fires during each time-step is a Poisson \emph{random variate}, with known parameter~\citep{__ref__c646c0890ccf4a59835f2df5c824c223}. By repeatedly advancing by a time-step of length $\tau$ through time\footnote{Here, and throughout the rest of the manuscript, we will assume that $\tau$ divides $T$ exactly.}, a sample path can be generated. An implementation of the tau-leap method is provided in Algorithm \ref{__label__8c11b055698d4dde8c58bd69ebc34f71}: to emphasise that the dynamics of a tau-leap sample path are different to the dynamics of the process $\boldsymbol{X}$, in our algorithm we denote the population of a tau-leap sample path as $\boldsymbol{Z}$.

\begin{algorithm}[ht]
\caption{The tau-leap method. This simulates a sample path using fixed time-step $\tau$.\protect\vphantom{$A_A^A$}}
\label{__label__8c11b055698d4dde8c58bd69ebc34f71}
 \begin{algorithmic}[1]
  \Require initial conditions, $\bs{Z}(0)$, time-step $\tau$, and terminal time, $T$.\protect\vphantom{$A_A^A$}
  \State set $\bs{Z} \leftarrow \bs{Z}(0)$ and set $t \leftarrow 0$
  
  \While{$t < T$}
  \For {each $R_j$}
  \State calculate propensity value $p_j(\bs{Z})$
  \State generate $K_j \sim \mathcal{P}(p_j(\bs{Z}) \cdot \tau)$
  \EndFor 
  \State set $\bs{Z} \leftarrow \bs{Z} + \sum_{j=1}^M K_j \cdot \bs{\nu}_j$
  \State set $t \leftarrow t + \tau$
  \EndWhile
 \end{algorithmic}
\end{algorithm}

Under reasonably general circumstances, it can be shown that the tau-leap method can generate sample paths more quickly then the DM~\citep{__ref__3e0b45aec9df4e05b2f3e8165dab4878}. However, if we choose to estimate a summary statistic with an ensemble of sample paths generated using the tau-leap method, then the resultant estimate will be biased. The bias is a consequence of using approximate reaction propensities in Algorithm \ref{__label__8c11b055698d4dde8c58bd69ebc34f71}; the bias typically scales as $\mathcal{O}(\tau)$~\citep{__ref__d5682854157f4e11b42663588059cb23}. Roughly speaking, the total CPU time scales with the number of time-steps, i.e. $\mathcal{O}(1/\tau)$. Thus, $\tau$ must be chosen to balance the competing demands of speed and accuracy. Having explained how one might implement the tau-leap method, we turn to discussing its mathematical representation.

\subsection{Representing the tau-leap method using Poisson processes}
In this section, we use the RTCR to motivate and represent the tau-leap method. We recall that the state $\bs{X}$ of our biochemical reaction network evolves according to Equation \eqref{__label__0ec1364809bb422f9cb945e0db5d4701}, which we now restate: \begin{equation*}
\bs{X}(T) = \bs{X}(0) + \sum_{j=1}^{M}  \mathcal{Y}_j \left(0,\int_0^T p_j(\bs{X}(t)) \wrt{t}\right) \cdot \nu_j. 
\end{equation*} We continue to represent the state of a tau-leap process at time $t$ as $\bs{Z}(t)$. Following from Algorithm \ref{__label__8c11b055698d4dde8c58bd69ebc34f71}, we suppose that $[0,T]$ is divided into $K$ equal time-steps of length $\tau$. The tau-leap assumption is that the propensities can change only at fixed times $t= k\cdot\tau$ (for $k = 1, 2, \dots$). Therefore, when we implement the tau-leap method we use the following approximation: \begin{equation}
\int_0^T p_j(\bs{Z}(t)) \wrt{t} \approx \sum_{k=0}^{K-1} p_j(\bs{Z}(\tau \cdot k))\cdot \tau. \label{__label__d75eb719ffeb4a1785990be711eb82b7}
\end{equation} If we insert assumption \eqref{__label__d75eb719ffeb4a1785990be711eb82b7} into Equation \eqref{__label__0ec1364809bb422f9cb945e0db5d4701}, then the evolution of the state of the tau-leap process, $\bs{Z}(t)$, is described by \begin{equation} 
\bs{Z}(T) =  \bs{Z}(0) + \sum_{j=1}^M \mathcal{Y}_j \left(0,  \sum_{k=0}^{K-1} {p_j}(\bs{Z}(\tau \cdot k))\cdot \tau  \right) \cdot \bs{\nu}_j, \label{__label__147b3b5568b84b7fb9350d16407a2ed2}
\end{equation} where the $\mathcal{Y}_j$ ($j = 1, \dots, M$) are unit-rate Poisson processes. We now rearrange Equation \eqref{__label__147b3b5568b84b7fb9350d16407a2ed2} so that it is easier to work with. First we set\footnote{Note that $j$ indexes the reaction, $R_j$; and $k$ indexes the time.} \begin{equation}
P_{k,j} = \sum_{k'=0}^{k} {p_j}(\bs{Z}(\tau \cdot k'))\cdot \tau, \label{__label__5be22b8f504c411085a039b28e63a9be}
\end{equation} with the special case of $P_{-1,j} = 0$. Then, we re-arrange Equation \eqref{__label__147b3b5568b84b7fb9350d16407a2ed2} to give \begin{equation}
\bs{Z}(T) =  \bs{Z}(0) + \sum_{k=0}^{K-1} \sum_{j=1}^M \mathcal{Y}_j \left(P_{k-1,j}, P_{k,j}  \right) \cdot \bs{\nu}_j. \label{__label__ef82539d7e0d4432b4309c17cf3c09c7}
\end{equation} 

The tau-leap method can be seen as a method for iterating over $k$: for each reaction $R_j$, at each step we calculate the number of events in the Poisson process $\mathcal{Y}_j$ between positions (or internal times) $P_{k-1, j}$ and $P_{k, j}$ (given by $\mathcal{Y}_j(P_{k-1, j}, P_{k, j})$).  Equation \eqref{__label__ef82539d7e0d4432b4309c17cf3c09c7} can therefore be re-arranged into an update formula. For $k = 1, \dots, K$, \begin{equation}
\bs{Z}(k\cdot\tau) =  \bs{Z}((k-1)\cdot\tau) + \sum_{j=1}^M \mathcal{Y}_j \left(P_{k-1,j}, P_{k,j}  \right) \cdot \bs{\nu}_j. \label{__label__42123ab026d8484d80c8be913d2804f8}
\end{equation}

\subsection{The R-leap method} \label{__label__fe56d15596ea45beb4d02983da69e032}
The R-leap method is an example of a simulation algorithm that generates approximate sample paths by firing a user-specified number of reactions during each step. The R-leap method was described by \citet{__ref__d45dab7799794c30abe917c6253324aa}. The R-leap method differs from the tau-leap method in that each `tau-leap' covers a fixed time interval (with the number of reaction events determined by simulation), whereas the time interval covered by each `R-leap' must be determined by simulating a random number (though the number of reaction events is fixed).  

The algorithm proceeds as follows. The user specifies how many reactions will fire during each simulation step: this quantity is labelled as $\mathcal{K}$. The algorithm starts at time $t = 0$, and the state vector, $\boldsymbol{Z}$, is used to describe the molecular populations. At each step, the time for the $\mathcal{K}$ reactions to take place, $\Delta$, is generated. Then, the precise combination of reactions that take place (e.g. three $R_1$ reactions, four $R_2$ reactions, and so on) is chosen. Finally, the state vector, $\boldsymbol{Z}$, and time, $t$, are updated. These steps are repeated as required. We now discuss the details of the algorithm. 

In the Gillespie DM (see Algorithm \ref{__label__c0331a4be21941daa64b9da1f158abd1}), the waiting time until the next reaction is given by an exponential variate with rate $p_0 = \sum_{j=1}^M p_j$. Consequently, given the fixed propensity values, the total waiting time for $\mathcal{K}$ reactions to fire in the R-leap method algorithm, $\Delta$, is given by the sum of $\mathcal{K}$ exponential variates\footnote{As the propensities are only updated every $\mathcal{K}$ reactions, $p_0$ is fixed for this many reactions.}. This sum is Gamma distributed, i.e. $\Delta \sim \Gamma(\mathcal{K}, 1/p_0)$, with shape parameter $\mathcal{K}$ and scale parameter $1/p_0$. 

Having calculated the distribution of the time period over which the next $\mathcal{K}$ reaction events occur, we must decide on the specific combination of reaction types that take place. This can be achieved with a conditional binomial method that proceeds as follows. For each of the $\mathcal{K}$ reaction events that we will simulate, the probability that it is a reaction of type $R_j$ is given by $p_j / p_0$. We start by considering $R_1$. A binomial random number, $\mathcal{B}(\mathcal{K}, p_1/p_0)$ is simulated, using $\mathcal{K}$ trials, each with probability $p_1/p_0$ of success. This provides the number of reactions of type $R_1$; we label this quantity as $K_1$. There are therefore $\mathcal{K} - K_1$ reactions that still need to be assigned to a type, and we know that these reactions are \emph{not} $R_1$ reactions. Thus, we generate a binomial random number, $K_2 \sim \mathcal{B}(\mathcal{K} - K_1, p_2 / (p_0 - p_1))$ to decide how many reactions of type $R_2$ have fired. In this case, there are $\mathcal{K} - K_1$ trials, each with conditional probability $p_2 / (p_0 - p_1)$ of success. The number of reactions of type $R_3$ is given by $K_3 \sim \mathcal{B}(\mathcal{K} - K_1 - K_2, p_3/(p_0 - p_1 - p_2))$. This process repeated until all $\mathcal{K}$ reactions have been assigned a type. To ensure that the final step is carried out properly, so that the sample path terminates at time $t = T$, we require the following rule: 

\textbf{Rule 1}. Conditioned on the value of $\Delta$, where $\Delta$ is the waiting time for $\mathcal{K}$ events to fire, the number of events that fire after traversing $\Delta'$ units of time, where $\Delta' < \Delta$, will be given by \begin{equation}
\mathcal{K}' = \mathcal{B}(\mathcal{K} - 1, \Delta' / \Delta).
\end{equation}  This follows as $\mathcal{K} - 1$ events take place strictly inside the interval of length $\Delta$, with these events uniformly distributed throughout the interval.  

A pseudo-code implementation of the R-leap method is provided in Algorithm \ref{__label__a399905995f94d32a3f5d4befd66bb4d}. Note that, when $\mathcal{K}$ decreases, the accuracy of the R-leap method increases, and when $\mathcal{K} = 1$, the DM is recovered.  The computational cost scales as $\mathcal{O}(\mathcal{K}^{-1})$. 

\begin{algorithm}[htb]
\caption{The R-leap method. This simulates a single sample path using fixed jump size, $\mathcal{K}$.\protect\vphantom{$A_A^A$}}
\label{__label__a399905995f94d32a3f5d4befd66bb4d}
 \begin{algorithmic}[1]
  \Require initial conditions, $\bs{Z}(0)$, jump size, $\mathcal{K}$, and terminal time, $T$.\protect\vphantom{$A_A^A$}
  \State set $\bs{Z} \leftarrow \bs{Z}(0)$ and set $t \leftarrow 0$
  
  \While{$t < T$}
  \State for each $R_j$, calculate propensity value $p_j(\bs{Z})$, and set $p_0 \leftarrow \sum_{j=1}^M p_j$
  \State generate $\Delta \sim \Gamma(\mathcal{K}, 1/p_0)$
  \If{$t + \Delta > T$}
  \State generate $\mathcal{K} \sim \mathcal{B}(\mathcal{K} - 1, (T-t) / \Delta)$, and set $t \leftarrow T$
  \Else
  \State set $t \leftarrow t + \Delta$
  \EndIf
  \For {$j = 1, \dots, M$}
  \State generate $K_j \sim \mathcal{B}\big(\mathcal{K}, p_j/\sum_{j'=j}^Mp_{j'}\big)$ and set $\mathcal{K} \leftarrow \mathcal{K} - K_j$
  \EndFor 
  \State set $\bs{Z} \leftarrow \bs{Z} + \sum_{j=1}^M K_j \cdot \bs{\nu}_j$
  \EndWhile
 \end{algorithmic}
\end{algorithm}

\subsection{Outlook}

In the next section, we will introduce the multi-level method. Our investigation of the multi-level method is motivated by the dramatic computational savings it leads to: compared with traditional simulation methods, the multi-level method can potentially reduce the CPU time required to estimate a summary statistic by orders of magnitude~\citep{__ref__b7cd69d78ffc4c2ba54f2f3e26b5d981}.

\section{Multi-level Monte Carlo} \label{__label__471d7eaf487749a0885bc893309e52bd}

In 2008, \citet{__ref__dbd754638429454ab66f23e25ec8f0be} described and implemented the multi-level framework; an efficient Monte Carlo scheme for stochastic differential equations. The multi-level method was subsequently extended to efficiently estimate summary statistics of discrete-state, continuous time Markov chains by \citet{__ref__5f82fa75a5314b0484ebeb52de9e2750}. The important contribution by \citet{__ref__5f82fa75a5314b0484ebeb52de9e2750} provides one way in which the multi-level technique can be used to model discrete-state systems, but there are many alternative ways in which the multi-level technique can be implemented. In this work, we will consider some of these. 

In an effort to increase computational efficiency, the multi-level method divides the work done in calculating a summary statistic of interest into parts, known as \emph{levels}. By way of example, the statistic we are interested in might be the average population of species $S_i$ at time $T$. We write the chosen summary statistic as  \begin{equation}
\mathcal{Q} = \mathbb{E}\left[ f(\boldsymbol{X})\right]. \label{__label__1ffd0e324b8d411185a134ba49a124a8}
\end{equation} Our aim is to estimate $\mathcal{Q}$ to within a given statistical accuracy as efficiently as possible. For each level $\ell = 0, 1, \dots, L, L+1$ (with $L$ specified later), we will define an estimator $\mathcal{Q}_\ell$ such that \begin{equation}
\mathcal{Q} = \sum_{\ell = 0}^{L+1} \mathcal{Q}_\ell. \label{__label__f5d42d8aaccd45a1b2049c1ba38f9fd0}
\end{equation} Each value of $\ell$ represents a level. We will independently estimate each $\mathcal{Q}_\ell$, and then sum our estimates up, so that we have an estimated value of $\mathcal{Q}$. As outlined in Section \ref{__label__6fd124019bec45a8bc6d5c32ae94030a}, a confidence interval can be constructed around our estimated value of $\mathcal{Q}$. The size of the confidence interval depends on the estimator variance, with the estimator variance of $\mathcal{Q}$ equal to the sum of the estimator variances of the estimates for $\mathcal{Q}_\ell$, (for $\ell = 0, \dots, L+1$). The key savings provided by the multi-level method arise because each of the estimates of $\mathcal{Q}_1, \dots, \mathcal{Q}_{L+1}$ can be calculated using a variance reduction technique. For each level of accuracy, a variance reduction technique cleverly lowers the number of sample paths required to estimate a statistic of interest, which means that, with a given level of accuracy, Monte Carlo simulation can be completed quickly. We explain the principle of variance reduction as follows: 

\textbf{Variance reduction.} Suppose we wish to estimate a summary statistic of a stochastic process, $\phi$, given by $\mathbb{E}[f(\phi)]$, where $f(\cdot)$ is a suitable function. Further, we suppose that estimating $\mathbb{E}[f(\phi)]$ is computationally intensive, because $f(\phi)$ has a relatively high sample variance, $\mathcal{V}_{f(\phi)}$. If we estimate $\mathbb{E}[f(\phi)]$ using $\mathcal{N}$ sample paths, then the estimator variance is given by $\mathcal{V}_{f(\phi)} / \mathcal{N}$. 

If there is a different stochastic process, $\psi$, and function, $g(\cdot)$, such that \begin{equation}
\mathbb{E}[f(\phi)] = \mathbb{E}[g(\psi)],
\end{equation} then we can instead estimate $\mathbb{E}[g(\psi)]$, and any resultant estimate is also an estimate for $\mathbb{E}[f(\phi)]$. The bias of the estimate remains unchanged. However, the estimator variance of the estimate is now given by $\mathcal{V}_{g(\psi)} / \mathcal{N}$, where $\mathcal{V}_{g(\psi)}$ is the variance of $g(\psi)$. Thus, if $\mathcal{V}_{g(\psi)} < \mathcal{V}_{f(\phi)}$, fewer sample paths are required to achieve a given confidence in the estimator when $\mathbb{E}[g(\psi)]$, and not $\mathbb{E}[f(\phi)]$, is estimated. Therefore, the overall simulation time can be reduced by estimating $\mathbb{E}[g(\psi)]$ instead of $\mathbb{E}[f(\phi)]$. 

Returning to the multi-level method: if the sample paths used to estimate $\mathcal{Q}_\ell$ have variance $\mathcal{V}_\ell$, and there are $\mathcal{N}_\ell$ such sample paths, then estimator variance of $\mathcal{Q}_\ell$ is given by $\mathcal{V}_\ell / \mathcal{N}_\ell$. When we define $\mathcal{Q}_\ell$ below, we use the principle of variance reduction to choose $\mathcal{Q}_\ell$ so that $\mathcal{V}_\ell$ is small. This means that $\mathcal{N}_\ell$ does not need to be big, and the sample paths can be completed quickly.  

Of course, the estimators $\mathcal{Q}_0, \dots, \mathcal{Q}_{L+1}$ can be produced in many different ways. In this section, we will use the tau-leap method to construct the aforementioned estimators. We now describe the calculations performed on each of levels $\ell = 0, \dots, L + 1$. 

\textbf{The base level.} The first level, $\ell = 0$, is known as the base level. We will generate sample paths using the tau-leap method, with  constant time-step $\tau_0$. The simulation cost of each sample path is denoted $\mathcal{C}_0$. We will choose a large value for $\tau_0$, so that $\mathcal{C}_0$ is relatively small. We label the complete trajectory of each of the $\mathcal{N}_0$ sample paths we generate as $\boldsymbol{Z}_0^{(r)}$, where $r=1, \dots, \mathcal{N}_0$. Thus, we estimate $\mathcal{Q}_0$ as \begin{equation*}
 \mathcal{Q}_0 \approx \frac{1}{\mathcal{N}_0} \sum_{r = 1}^{\mathcal{N}_0} f\Big(\boldsymbol{Z}_0^{(r)}\Big).
\end{equation*} We expect that our estimate, $\mathcal{Q}_0$, will be highly biased due to the large choice of $\tau_0$.

\textbf{The intermediate correction levels.} Subsequent, intermediate levels, indexed as $\ell = 1, \dots, L$, are known as the correction levels. We consider pairs of sample paths, $(\boldsymbol{Z}_\ell, \boldsymbol{Z}_{\ell-1})$, where $\boldsymbol{Z}_\ell$ is simulated using the tau-leap method with a time-step $\tau_\ell$, and $\boldsymbol{Z}_{\ell-1}$ is also simulated using the tau-leap method, but with a time-step $\tau_{\ell-1}$. We follow \citet{__ref__5f82fa75a5314b0484ebeb52de9e2750} in choosing an integer refinement factor, $\mathcal{M}$, so that $\tau_\ell = \tau_{\ell-1} / \mathcal{M}$ (i.e. $\tau_\ell$ is $\mathcal{M}$ times smaller than  $\tau_{\ell-1}$). Then, our estimator for level $\ell$ is \begin{equation*}
 \mathcal{Q}_\ell \approx \frac{1}{\mathcal{N}_\ell} \sum_{r = 1}^{\mathcal{N}_\ell} \left[ f\Big(\boldsymbol{Z}_\ell^{(r)}\Big) - f\Big(\boldsymbol{Z}_{\ell-1}^{(r)}\Big) \right].
\end{equation*} The CPU cost of generating each pair of sample paths is denoted as $\mathcal{C}_\ell$. The stochastic simulation method that is used to estimate each sample value will be carried out with a variance reduction technique that ensures that the variance, $\mathcal{V}_\ell$, is small. In total there are $L$ intermediate levels that reduce the estimator bias. {The optimal value of $L$ can be chosen by following a systematic search procedure (reference \citep{__ref__b7cd69d78ffc4c2ba54f2f3e26b5d981} contains a detailed discussion).}

\textbf{The final level.} Optionally, the final level, $L+1$, removes all remaining bias\footnote{If a low-bias estimate can be tolerated, then this step can be skipped, and the overall CPU time is consequently reduced. {The resultant bias can be estimated by following the approach presented in \citet{__ref__dbd754638429454ab66f23e25ec8f0be}.}}. We compare pairs of sample paths, $(\boldsymbol{X}, \boldsymbol{Z}_{L})$, one of which is unbiased, and the other a tau-leap method sample path with time-step $\tau_L$. The estimator on this level is equal to \begin{equation*}
 \mathcal{Q}_{L+1} \approx \frac{1}{\mathcal{N}_{L+1}} \sum_{r = 1}^{\mathcal{N}_{L+1}} \left[ f\Big(\boldsymbol{X}^{(r)}\Big) - f\Big(\boldsymbol{Z}_{L}^{(r)}\Big) \right].\label{__label__ef9d0ea6699f42d2bc026e6e5e264db9}
\end{equation*} The CPU time required for each individual pair of sample paths is given by $\mathcal{C}_{L+1}$.

\subsection{Configuring the multi-level method}
We assess the efficiency of the multi-level method by determining the total CPU time required to estimate a summary statistic of interest. As each sample point on level $\ell$ takes $\mathcal{C}_\ell$ units of CPU time to generate, the total CPU time is given by $\sum_{\ell=0}^{L+1} \mathcal{C}_\ell \mathcal{N}_\ell$. Each level is independently simulated; therefore, the total estimator variance, $\mathcal{\widehat{V}}$, is given by \begin{equation} \mathcal{\widehat{V}} = \sum_{\ell = 0}^{L+1} \mathcal{V}_\ell / \mathcal{N}_\ell. \label{__label__b0d92eb1d30945e39fbbb5b435987319}\end{equation} If we wish to constrain the overall estimator to within a statistical error of $\varepsilon$, and the values of $\mathcal{V}_\ell$ are known, then it can be shown that that the total computational cost is minimized by choosing $\mathcal{N}_\ell$ (for $\ell = 0, 1, \dots, L+1$) to be~\citep{__ref__b7cd69d78ffc4c2ba54f2f3e26b5d981}  \begin{equation}\mathcal{N}_\ell = \left\{\frac{1}{\varepsilon}\sum_{m=0}^{L+1}\sqrt{\mathcal{V}_m \cdot \mathcal{C}_m}\right\} \sqrt{\frac{\mathcal{V}_\ell}{\mathcal{C}_\ell}}. \label{__label__d6427797c09740c6ac334b3f0847a63e} \end{equation} The values of $\mathcal{V}_\ell$ are, however, usually unknown, and must therefore be estimated. Typically, a small number of trial sample paths are generated, and $\mathcal{V}_\ell$ is thus estimated; we will discuss the merits of this approach in Section \ref{__label__439a9eb968c54826b8730e3976b4d497}. 
 
If Equation \eqref{__label__d6427797c09740c6ac334b3f0847a63e} is used to specify $\mathcal{N}_\ell$, then the total simulation cost is given by \begin{equation}
\mathcal{C} = \frac{1}{\varepsilon}  \left\{\sum_{\ell = 0}^{L+1} \sqrt{\mathcal{V}_\ell \cdot \mathcal{C}_\ell} \right\}^2. \label{__label__436d1497303048038259cf5f5c78ec9c}
\end{equation} Thus, it is to our advantage to ensure that the sample variance on each level, $\mathcal{V}_\ell$, is as small as possible. We outline the variance reduction technique employed by \citet{__ref__5f82fa75a5314b0484ebeb52de9e2750} to reduce the variance of the correction levels (i.e. where $\ell = 1, \dots, L$), and mention a second variance reduction technique for the final level (i.e. $\ell = L + 1$). 

We first explain how a variance reduction technique benefits the correction level estimators, $\mathcal{Q}_\ell$, for $\ell = 1, \dots, L$. To generate the $r$-th sample value of $\mathcal{Q}_\ell$, $\left[ f\Big(\boldsymbol{Z}_\ell^{(r)}\Big) - f\Big(\boldsymbol{Z}_{\ell-1}^{(r)}\Big) \right]$, we need to generate two sample paths using the tau-leap method, but with different time-steps. As we are constructing a Monte Carlo estimator, we require each of the sample values to be independent of the other sample values. The key point to note is that for each $r$, there is no need for $ f\Big(\boldsymbol{Z}_\ell^{(r)}\Big)$ and $ f\Big(\boldsymbol{Z}_{\ell-1}^{(r)}\Big)$ to be independent of one another. Recall that \begin{align*}
\mathcal{V}\left[ f\Big(\boldsymbol{Z}_\ell^{(r)}\Big) - f\Big(\boldsymbol{Z}_{\ell-1}^{(r)}\Big) \right] =  \phantom{a} &\mathcal{V} \left[ f\Big(\boldsymbol{Z}_\ell^{(r)}\Big)\right] + \mathcal{V}\left[f\Big(\boldsymbol{Z}_{\ell-1}^{(r)}\Big) \right]\\ & - 2\,\text{Cov} \left[ f\Big(\boldsymbol{Z}_\ell^{(r)}\Big), f\Big(\boldsymbol{Z}_{\ell-1}^{(r)}\Big) \right]. \end{align*} We note it is therefore in our interests for $\boldsymbol{Z}_\ell^{(r)}$ and $ \boldsymbol{Z}_{\ell-1}^{(r)}$ to exhibit a strong, positive correlation. This, in turn, should give rise to a lower estimator variance, and the total CPU time given by Equation \eqref{__label__436d1497303048038259cf5f5c78ec9c} is reduced. 

The final level, $\ell = L + 1$, involves coupling an unbiased sample path, $\boldsymbol{X}^{(r)}$, to a biased sample path, $\boldsymbol{Z}_L^{(r)}$. The estimator variance of samples of $\mathcal{Q}_{L+1}$ can be similarly reduced. Reference \citep{__ref__b7cd69d78ffc4c2ba54f2f3e26b5d981} contains additional information on how to do this.

In Sections \ref{__label__788405dd7a434bb5acdbf2a30ef377af} and \ref{__label__576de80bab724c3284173a8363d971e7}, we will discuss two distinct variance reduction approaches. The key idea is to use the same random input, as far as possible, for both $\boldsymbol{Z}_\ell$ and $\boldsymbol{Z}_{\ell -1}$. We explain the Split propensity method (SPM)~\citep{__ref__5f82fa75a5314b0484ebeb52de9e2750} in Section \ref{__label__788405dd7a434bb5acdbf2a30ef377af}, the previous gold-standard for multi-level Monte Carlo. The Common process method (CPM) is explained in Section \ref{__label__576de80bab724c3284173a8363d971e7}: this has been previously used to develop low-variance estimators for parameter sensitivity analysis~\citep{__ref__06e05ee3d1954ff58fd9cb7791b406de}, but, until now, has yet to be used for multi-level Monte Carlo simulation. Having outlined the multi-level method, we will compare the two different implementations. We assess each implementation for robustness and improvements in numerical performance. 

\section{Split propensity method} \label{__label__788405dd7a434bb5acdbf2a30ef377af}
Here we present the method developed by \citet{__ref__5f82fa75a5314b0484ebeb52de9e2750} that we call the SPM. The SPM involves the \emph{coupling} of two tau-leap processes that are required to provide sample paths $\boldsymbol{Z}_\ell$ and $\boldsymbol{Z}_{\ell -1}$ (we discuss $\mathcal{Q}_{L+1}$ separately below). For brevity, we will refer to a pair of sample paths as comprising a coarse path (referring to $\boldsymbol{Z}_{\ell-1}$, with $\tau = \tau_C = \tau_0 / \mathcal{M}^{\ell-1}$) and a fine path (referring to $\boldsymbol{Z}_\ell$, with $\tau = \tau_F = \tau_0 / \mathcal{M}^{\ell}$). The SPM is premised as follows: we aim to keep the $r$-th sample paths of the approximate processes with time-steps $\tau_C$ and $\tau_F$ as similar to each other as possible over the time period of interest. 
The SPM is implemented by creating `virtual reaction channels'. Each reaction, $R_j$, is split into three virtual channels, labelled as $R^1_j$, $R^2_j$ and $R^3_j$, defined such that:
\begin{itemize}
 \item $R^1_j$ : reaction $R_j$ fires in both the coarse and fine sample paths; 
 \item $R^2_j$ : reaction $R_j$ fires only in the coarse sample path;
\item $R^3_j$ : reaction $R_j$ fires only in the fine sample path.
\end{itemize} If $R_j$ has propensity $p_j^C$ in the coarse sample path, and $p_j^F$ in the fine sample path, then the propensities of $R^k_j$, for $k \in \{1, 2, 3\}$, are given by \begin{equation}
p^1_j = \min \big\{p^F_j, p^C_j\big\}, \hspace{5mm}  p^2_j = p^C_j - p^1_j, \hspace{5mm} p^3_j = p^F_j - p^{1}_j. \label{__label__8b0e5f6d7021402b8ab26991f440427d}
\end{equation} Note that at least one of $p^2_j$ and $p^3_j$ are zero. Then, if the state vectors $\boldsymbol{Z}_\ell$ and $\boldsymbol{Z}_{\ell -1}$ are similar, then we expect that, for each $j$, $p_j^C \sim p_j^F$, so that $p^1_j \gg p_j^2, p_j^3$. In this case, the majority of reactions will occur through reaction channels of the form $R_j^1$, so the state vectors $\boldsymbol{Z}_\ell$ and $\boldsymbol{Z}_{\ell -1}$ ought to remain somewhat similar after each time-step. Full pseudo-code is presented in Algorithm \ref{__label__e8a29791c9cc48bfa729f96cc24f48f6}.

\begin{algorithm}[hbt]
\caption{The SPM-coupled tau-leap method. This simulates a pair of sample paths.\protect\vphantom{$A_A^A$}}
\label{__label__e8a29791c9cc48bfa729f96cc24f48f6}
 \begin{algorithmic}[1]
  \Require initial conditions, $\bs{Z}(0)$, time-steps $\tau_C$, $\tau_F$, and terminal time, $T$.\protect\vphantom{$A_A^A$}
  \State set $\bs{Z}^C \leftarrow \bs{Z}(0)$, $\bs{Z}^F \leftarrow \bs{Z}(0)$ and $t \leftarrow 0$. Set \texttt{constant }$\mathcal{M} \leftarrow \tau_C / \tau_F$
  \For {each $t \in \{0, \tau_{\ell-1}, 2\cdot\tau_{\ell-1}, \cdots, T - \tau_{\ell-1}\}$}
  \State for each $R_j$, calculate propensity value $p^C_j(\bs{Z}^C)$
  \For{each $s \in \{t, t + \tau_{\ell}, \dots, t + (\mathcal{M}-1)\cdot\tau_{\ell}\}$}
  \State for each $R_j$,  calculate propensity value $p^F_j(\bs{Z}^F)$
  \State for each $R_j$, calculate virtual propensities $p^1_j$, $p^2_j$ and $p^3_j$ 
  \For{each $R_j$ and \textbf{for} each $k \in \{1,2,3\}$}
  \State generate $K_{jk} \sim \mathcal{P}(p^k_j \cdot \tau_{\ell})$
  \EndFor
  \State set $\bs{Z}^C \leftarrow \bs{Z}^C + \sum_{j=1}^M (K_{j1}+K_{j2}) \cdot \bs{\nu}_j$
  \State set $\bs{Z}^F \leftarrow \bs{Z}^F + \sum_{j=1}^M (K_{j1}+K_{j3}) \cdot \bs{\nu}_j$
  \EndFor
  \EndFor 
 \end{algorithmic}
\end{algorithm}

\subsection{Estimating $\mathcal{Q}_{L+1}$}
The final estimator, $\mathcal{Q}_{L+1}$, couples a tau-leap process $\boldsymbol{Z}_L$ (with time-step given by $\tau_L$), with an exact process, $\boldsymbol{X}$. The SPM procedure demonstrated above is followed to produce three virtual channels. The resultant system is simulated with a suitable algorithm; pseudo-code is provided in Algorithm \ref{__label__b2a50be6a1a34f8ea4a20567177a1f61}. 

\section{Common process method} \label{__label__576de80bab724c3284173a8363d971e7}
In this section, we present the CPM. As with the SPM, the CPM involves the \emph{coupling} of two tau-leap processes that are required to provide sample paths for $\boldsymbol{Z}_\ell$ and $\boldsymbol{Z}_{\ell-1}$, so that we can efficiently estimate $\mathcal{Q}_\ell$, where $\ell = 1, \dots, L$. As before, we will refer to a pair of sample paths as comprising a coarse (referring to $\boldsymbol{Z}_{\ell-1}$) and a fine (referring to $\boldsymbol{Z}_\ell$) path.

\subsection{Estimating $\mathcal{Q}_\ell$, where $\ell = 1, \dots, L+1$}
We now discuss the CPM, and use the RTCR provided by Equation \eqref{__label__ef82539d7e0d4432b4309c17cf3c09c7} to represent the processes $\boldsymbol{Z}_\ell$ (with time-step $\tau_\ell$) and $\boldsymbol{Z}_{\ell-1}$ (with time-step $\tau_{\ell-1}$): \begin{align*}
&\boldsymbol{Z}_\ell(T) =  \boldsymbol{Z}_\ell(0) + \sum_{k=0}^{K} \sum_{j=1}^M \mathcal{Y}_j \left(P_{\ell,k-1, j}, P_{\ell,k, j}  \right) \cdot \boldsymbol{\nu}_j; \\
&\boldsymbol{Z}_{\ell-1}(T) =  \boldsymbol{Z}_{\ell-1}(0) + \sum_{k=0}^{K'} \sum_{j=1}^M \mathcal{Y}_j \left(P_{\ell-1,k-1, j}, P_{\ell-1,k, j}  \right) \cdot \boldsymbol{\nu}_j,
\end{align*} where \begin{equation}
P_{\ell,k, j} = \sum_{k'=0}^{k} {p_j}(\boldsymbol{Z}_\ell(\tau_\ell \cdot k'))\cdot \tau_\ell, 
\end{equation} and the $\mathcal{Y}_j$ ($j = 1, \dots, M$) are unit-rate Poisson processes. The CPM method can produce a low variance estimate for $\mathcal{Q}_\ell$ by using the same set of $M$ Poisson processes, i.e. $\mathcal{Y}_1, \dots, \mathcal{Y}_M$, for both sample paths $\boldsymbol{Z}_\ell$ and $\boldsymbol{Z}_{\ell-1}$. 

This CPM scheme can be implemented by essentially running the tau-leap algorithm twice. We explain the procedure in detail in the following paragraphs, but the method can be summarised as follows: \begin{itemize}
\item during the first phase, we simulate a sample path $\boldsymbol{Z}_\ell$ with time-step $\tau_{\ell}$. In doing so, the total number of times each Poisson process, $\mathcal{Y}_j$ (for $j = 1, \dots, M$), has fired over each time-step is stored into memory (see Section \ref{__label__314d91af47254ba2970e65b7a523deea});
\item during the second phase, we simulate a sample path $\boldsymbol{Z}_{\ell-1}$, making use of the Poisson processes stored in memory and using interpolation (as phases one and two will use distinct time-steps) as required (see Section \ref{__label__d9e3c1a0a64042d3ad74b594801a59cf}). 
\end{itemize} A sample path for the process $\boldsymbol{Z}_{\ell}$ will be simulated according to the pseudo-code provided in Algorithm \ref{__label__a213e7f2d83d43b5bda17ba4873838cc}; a sample path for process $\boldsymbol{Z}_{\ell-1}$ will be simulated according to Algorithm \ref{__label__3f018c85cadc4466bc6243321e791d24}.

{The CPM has been adapted from a technique used to produce low variance estimates for parametric sensitivities~\citep{__ref__06e05ee3d1954ff58fd9cb7791b406de}. Compared with the simulation methods presented in \citet{__ref__06e05ee3d1954ff58fd9cb7791b406de}, this approach differs in two important ways. Firstly, following the tau-leap assumption, we generate pairs of sample paths by firing multiple reactions at once, whereas \citet{__ref__06e05ee3d1954ff58fd9cb7791b406de} fire reactions singly. Secondly, \citet{__ref__06e05ee3d1954ff58fd9cb7791b406de} generate pairs of sample paths with slightly different propensity values, whereas we generate pairs of sample paths with different time-steps.}

\begin{algorithm}[bth]
\caption{\protect\vphantom{$A^A$}This simulates a pair of SPM-coupled sample paths: an exact path and a path with the tau-leap method using time-step $\tau$.\protect\vphantom{$A_A$}}
\label{__label__b2a50be6a1a34f8ea4a20567177a1f61}
 \begin{algorithmic}[1]
  \Require initial conditions, $\bs{X}(0)$, time-step, $\tau$, and terminal time, $T$.\protect\vphantom{$A_A^A$}
  \State set $\bs{X} \leftarrow \bs{X}(0)$, $\bs{Z} \leftarrow \bs{X}(0)$, $t \leftarrow 0$, and $t^* \leftarrow \tau$
  \State for each $R_j$, calculate propensity values $p^X_j(\bs{X})$ and $p^Z_j(\bs{Z})$
  \Loop
  \State for each $R_j$, set $p^*_j = \max\big\{p^X_j, p^Z_j\big\}$, and set $p_0^* = \sum_{j=1}^M p_j^*$
  \State set $\Delta \leftarrow \text{Exp}(p^*_0)$  
  \If{$t + \Delta > T$}
  \State \textbf{break}
  \ElsIf {$t + \Delta > t^*$}
  \State set $t \leftarrow t^*$, and $t^* \leftarrow t^* + \tau$
  \State for each $R_j$, calculate propensity values $p^Z_j(\bs{Z})$ 
  \Else
  \State choose index $k$, where $k$ has probability $p^*_k / p^*_0$ of being chosen
  \State with probability $p^X_k / p^*_k$, set $\bs{X} \leftarrow \bs{X} + \bs{\nu}_k$
  \State with probability $p^Z_k / p^*_k$, set $\bs{Z} \leftarrow \bs{Z} + \bs{\nu}_k$ 
  \State set $t \leftarrow t + \Delta$
  \State for each $R_j$, calculate propensity values $p^X_j(\bs{X})$
  \EndIf
  \EndLoop
 \end{algorithmic}
\end{algorithm}

\subsubsection{Phase one: simulating $\boldsymbol{Z}_\ell$} \label{__label__314d91af47254ba2970e65b7a523deea}
We first explain how to simulate a sample path of the tau-leap process $\boldsymbol{Z}_\ell$. We use the RTCR as a starting point. At time $t = 0$, the population is equal to the initial condition, $\boldsymbol{Z}(0)$. The populations at later times, $t = k\cdot \tau_\ell$ (for $k=1, 2, \dots$), are given by $\boldsymbol{Z}(k \cdot \tau_\ell)$. By setting $k = 1, 2, \dots$, the value of $\boldsymbol{Z}(k \cdot \tau_\ell)$ can be recursively calculated according to Equation \eqref{__label__147b3b5568b84b7fb9350d16407a2ed2}, that is \begin{equation*}
\boldsymbol{Z}_\ell(k\cdot \tau_\ell) =  \boldsymbol{Z}_\ell((k-1)\cdot \tau_\ell) + \sum_{j=1}^M \mathcal{Y}_j \left(P_{\ell,k-1, j}, P_{\ell,k, j}  \right) \cdot \boldsymbol{\nu}_j. \label{__label__c9ce18c4323a42f887608e24995560dc}
\end{equation*} Recall that the quantity $ \mathcal{Y}_j \left(P_{\ell,k-1, j}, P_{\ell,k, j}  \right)$ represents the number of arrivals of the unit-rate Poisson process $\mathcal{Y}_j$ over the interval $(P_{\ell,k-1, j}, P_{\ell,k, j}]$. The value of $ \mathcal{Y}_j \left(P_{\ell,k-1, j}, P_{\ell,k, j}  \right)$ is Poisson distributed with parameter $P_{\ell,k, j} - P_{\ell,k-1, j}$. Note that, in accordance with Equation \eqref{__label__5be22b8f504c411085a039b28e63a9be}, \begin{equation}
P_{\ell,k, j} - P_{\ell,k-1, j} = p_j(\boldsymbol{Z}_\ell(k \cdot \tau_\ell)) \cdot \tau_\ell.
\end{equation}
The number of times reaction $R_j$ occurs over the time span $((k-1)\cdot \tau_\ell, k\cdot \tau_\ell]$ is given by the Poisson random variate\footnote{Note that, as expected, this quantity is consistent with Algorithm \ref{__label__8c11b055698d4dde8c58bd69ebc34f71}.}, $\mathcal{P}(p_j(\boldsymbol{Z}_\ell(k \cdot \tau_\ell)) \cdot \tau_\ell)$. 

The tau-leap method provided in Algorithm \ref{__label__a213e7f2d83d43b5bda17ba4873838cc} is therefore implemented. At each step of the algorithm ($k = 1, 2, \dots)$, we will store the tuple $\langle p_j(\boldsymbol{Z}_\ell(k \cdot \tau_\ell)) \cdot \tau, K_j \rangle$, where $K_j$ describes the number of times reaction $R_j$ occurs over the time span. The tuple is stored in an ordered list $\mathcal{F}_j$. {If more time-steps are used, the size of each ordered list, $\mathcal{F}_j$, increases.}

The ordered list $\mathcal{F}_j$ therefore stores details of the arrivals of the unit-rate Poisson process $\mathcal{Y}_j$.

\textbf{Example.} Suppose that $\mathcal{F}_j = \left\{ \langle 2.1, 3 \rangle, \langle 4.0,7 \rangle , \langle 1.7 , 3 \rangle, \dots  \right\}.$ This means that we know that over the interval (of the unit-rate process) $(0.0, 2.1]$, there were three arrivals in the unit-rate Poisson process $\mathcal{Y}_j$. We also know that there were seven arrivals over the interval $(2.1, 6.1]$. Thus, over the interval $(0.0, 6.2]$, a total of ten arrivals were observed. 

Therefore, the ordered lists, $\mathcal{F}_j$ (for $j = 1, \dots, M$), provide \emph{partial information} about the arrivals of the unit-rate Poisson process $\mathcal{Y}_j$. Each ordered list, $\mathcal{F}_j$, contains information about the total number of arrivals between specific positions, but does \emph{not} contain information about the precise time at which each arrival is observed. Additional details will therefore be generated when they are needed for the coarse path. In the next section, we show how to use these ordered lists to generate a sample path of the process $\boldsymbol{Z}_{\ell-1}$ according to the CPM. 

\begin{algorithm}[htb]
\caption{\protect\vphantom{$A^A$}Phase one of the CPM-coupled tau-leap method. This simulates a single fine sample path, and records partial details of the Poisson processes associated with each reaction channel.\protect\vphantom{$A_A$}}
\label{__label__a213e7f2d83d43b5bda17ba4873838cc}
 \begin{algorithmic}[1]
  \Require initial conditions, $\bs{Z}(0)$, time-step, $\tau$, and terminal time, $T$.\protect\vphantom{$A_A^A$}
  \State set $\bs{Z} \leftarrow \bs{Z}(0)$ and set $t \leftarrow 0$
  
  \While{$t < T$}
  \For {each $R_j$}
  \State calculate propensity value $p_j(\bs{Z})$
  \State generate $K_j \sim \mathcal{P}(p_j(\bs{Z}) \cdot \tau)$
  \State add the tuple $\langle p_j(\bs{Z}) \cdot \tau, K_j \rangle$ to the end of ordered list $\mathcal{F}_j$
  \EndFor
  \State set $\bs{Z} \leftarrow \bs{Z} + \sum_{j=1}^M K_j \cdot \bs{\nu}_j$ and set $t \leftarrow t + \tau$
  \EndWhile
 \end{algorithmic}
\end{algorithm}

\subsubsection{Phase two: simulating $\boldsymbol{Z}_{\ell-1}$} \label{__label__d9e3c1a0a64042d3ad74b594801a59cf}
The CPM method uses the same unit-rate Poisson process, $\mathcal{Y}_j$, to fire the $R_j$ reactions in each of $\boldsymbol{Z}_{\ell}$ and $\boldsymbol{Z}_{\ell-1}$. Therefore, in this section we describe how to generate a sample path of process $\boldsymbol{Z}_{\ell-1}$ using the information stored in the ordered lists, $\mathcal{F}_j$ (for $j = 1, \dots, M$). As summarised above, the ordered list, $\mathcal{F}_j$, contains only an outline of the Poisson process, $\mathcal{Y}_j$ (for $j = 1, \dots, M$), and further details of these Poisson processes need to be filled in as required. The population of $\boldsymbol{Z}_{\ell-1}$ is determined by setting $k = 1, 2, \dots$, and recursively calculating \begin{equation*}
\boldsymbol{Z}_{\ell-1}(k\cdot \tau_{\ell-1}) =  \boldsymbol{Z}_{\ell-1}((k-1)\cdot \tau_{\ell-1}) + \sum_{j=1}^M \mathcal{Y}_j \left(P_{{\ell-1},k-1, j}, P_{{\ell-1},k, j}  \right) \cdot \boldsymbol{\nu}_j.
\end{equation*}{Note that, in accordance with Equation \eqref{__label__5be22b8f504c411085a039b28e63a9be}, \begin{equation}
P_{\ell-1,k, j} - P_{\ell-1,k-1, j} = p_j(\boldsymbol{Z}_{\ell-1}(k \cdot \tau_{\ell-1})) \cdot \tau_{\ell-1}.
\end{equation} Therefore, the simulation algorithms described in phases one and two are both mathematically equivalent to the tau-leap method (Algorithm \ref{__label__8c11b055698d4dde8c58bd69ebc34f71}). Thus, Algorithms \ref{__label__a213e7f2d83d43b5bda17ba4873838cc} and \ref{__label__3f018c85cadc4466bc6243321e791d24} are mathematically identical, and therefore, the sample paths that they produce are statistically indistinguishable.} From the ordered list $\mathcal{F}_j$ we can directly read off the number of arrivals of the Poisson process $\mathcal{Y}_j$ at positions $P_{l-1,0,j}, P_{l-1,1,j}, P_{l-1,2,j}, \dots$ (for $j =1, \dots, M$). At other positions, interpolation will be required. We use the following two rules to interpolate the Poisson process: 

\textbf{Rule 2}. If there are $K$ arrivals over the interval $(\alpha, \gamma)$ then, for $\beta$ between $\alpha$ and $\gamma$,  the number of arrivals over the interval $(\alpha, \beta)$ is binomially distributed\footnote{This follows as the $K$ arrivals are uniformly distributed over $(\alpha, \gamma)$.} \begin{equation}
\mathcal{B} \left(K, \frac{\beta - \alpha}{\gamma - \alpha}\right). 
\end{equation}

\textbf{Rule 3 (a caveat)}. If there are $K_1$ arrivals over the interval $(\alpha, \gamma)$, and $K_2$ arrivals over the interval $(\gamma, \beta)$, then interpolation using Rule 2 must be individually performed on the intervals $(\alpha, \gamma)$ and $(\gamma, \beta)$. Rule 2 cannot be directly applied to the entire interval $(\alpha, \beta)$ with $K_1 + K_2$ arrivals. 

Rules 2 and 3 provide all the tools needed for Monte Carlo simulation. We first illustrate interpolation of a Poisson process with an example; pseudo-code is then provided in Algorithm \ref{__label__3f018c85cadc4466bc6243321e791d24}. 

\begin{algorithm}[!htb]
\caption{\protect\vphantom{$A^A$}Phase two of the CPM-coupled tau-leap method. This simulates a single coarse sample path from the Poisson processes stored during phase one.\protect\vphantom{$A_A$}}

\label{__label__3f018c85cadc4466bc6243321e791d24}
 \begin{algorithmic}[1]
  \Require initial conditions, $\bs{Z}(0)$, time-step, $\tau$, ordered lists, $\mathcal{F}_j$ (with $j=1, \dots, M$), and terminal time, $T$.\protect\vphantom{$A_A^A$}
  \State set $\bs{Z} \leftarrow \bs{Z}(0)$ and set $t \leftarrow 0$
  
  \While{$t < T$}
  \For {each $R_j$}
  \State calculate propensity value $p_j(\bs{Z})$
  \State set $P \leftarrow 0$, $K \leftarrow 0$
  \While {$P < p_j(\bs{Z})\cdot \tau$}
  \If{$\mathcal{F}_j = \emptyset$}
  \State \textbf{break}
  \Else
  \State read and then delete $(P', K')$ from the front of $\mathcal{F}_j$
  \State set $P \leftarrow P + P'$ and $K \leftarrow K + K'$
  \EndIf
  \EndWhile
  \If{$P > p_j(\bs{Z}) \cdot \tau$}
  \State generate $K' \sim \mathcal{B}(K', (P - {p_j(\bs{Z})\cdot \tau})/{P'})$
  \State set $K_j \leftarrow K - K'$
  \State add the tuple $\langle P - p_j(\bs{Z}) \cdot \tau, K' \rangle$ to the front of ordered list $\mathcal{F}_j$
  \ElsIf {$P < p_j(\bs{Z}) \cdot \tau$}
  \State generate $K' \sim \mathcal{P}({p_j(\bs{Z})\cdot \tau}-P)$
  \State set $K_j \leftarrow K + K'$
  \Else
  \State set $K_j \leftarrow K$
  \EndIf
  \EndFor
  \State set $\bs{Z} \leftarrow \bs{Z} + \sum_{j=1}^M K_j \cdot \bs{\nu}_j$ and $t \leftarrow t + \tau$
  
  \EndWhile
 \end{algorithmic}
\end{algorithm}

\textbf{Example.} We return to our earlier example of an ordered list, $\mathcal{F}_j = \{ \langle 2.1, 3 \rangle, \langle 4.0, 7 \rangle , \langle 1.7 , 3 \rangle,$ $\dots$  $\}.$ Suppose we wish to determine $\mathcal{Y}_j (0.0, 5.0)$, the number of arrivals in the unit-rate Poisson process by time $5.0$. We first apply Rule 3: there are three arrivals over the time span $(0.0, 2.1]$, and we need to determine how many further arrivals are observed over the time span $(2.1, 5.0]$. We apply Rule 2 as follows: over the interval $(2.1, 6.1]$ there are seven arrivals, and we want to know know many of these arrivals occur inside the sub-interval $(2.1, 5.0)$. A binomial variate, $\mathcal{B}(7,2.9/4)$ is generated to interpolate the Poisson process.

\subsection{Estimating $\mathcal{Q}_{L+1}$}
The final estimator, $\mathcal{Q}_{L+1}$, couples a tau-leap process $\boldsymbol{Z}_L$ (with time-step given by $\tau_L$), with an exact process, $\boldsymbol{X}$. The CPM is implemented as outlined in the following paragraph. 

\begin{algorithm}[htb]
\caption{\protect\vphantom{$A^A$}Phase one of the CPM-coupled DM and tau-leap method. This simulates a single, exact sample path, and records the Poisson processes associated with each reaction channel.\protect\vphantom{$A_A$}}
\label{__label__4f095ab3883a4ddfb2954e2fa11f65f9}
\begin{algorithmic}[1]
  \Require initial conditions, $\bs{X}(0)$, and terminal time, $T$.\protect\vphantom{$A_A^A$}
  \State set $\bs{X} \leftarrow \bs{X}(0)$, and set $t \leftarrow 0$ 
  \State for each $R_j$, set $P_j \leftarrow 0$, generate $T_j \leftarrow \text{Exp}(1)$, and store $T_j$ as the first element of $\mathcal{F}_j$
  \Loop
  \State for each $R_j$, calculate propensity values $p_j(\bs{X})$ and calculate $\Delta_j$ as \begin{equation*} \Delta_j = \frac{T_j - P_j}{p_j} \end{equation*}
  \State set $\Delta \leftarrow \min_j \Delta_j$, and $k \leftarrow \text{argmin}_j \Delta_j$
  \If{$t + \Delta > T$}
  \State \textbf{break}
  \EndIf
  \State set $\bs{X}(t + \Delta) \leftarrow \bs{X}(t) + \bs{\nu}_j$, set $t \leftarrow t + \Delta$, and for each $R_j$, set $P_j \leftarrow P_j + p_j \cdot \Delta$
  \State generate $u \sim \text{Exp}(1)$, then set $T_k \leftarrow T_k + u$ and append $u$ to end of $\mathcal{F}_k$
  \EndLoop
\end{algorithmic}
\end{algorithm} 

To simulate a sample path of the exact process, $\boldsymbol{X}$, every reaction must be individually simulated. Therefore, we will need to determine every required arrival time of each Poisson process, $\mathcal{Y}_j$. The arrival times of each Poisson process are then saved into memory; the same set of Poisson processes is used to generate a sample path for $\boldsymbol{Z}_\ell$, the tau-leap process with time-step $\tau_L$. A range of algorithms can be satisfactorily implemented to simulate the sample paths. We will adapt the Modified Next Reaction Method (MNRM), as described by \citet{__ref__54e4fec61bb749dfacb9f7a516266af0}. The CPM proceeds in two phases, that can be outlined as follows: \begin{itemize}
\item during the first phase, a sample path for process $\boldsymbol{X}$ is generated using the MNRM. The waiting times of Poisson process $\mathcal{Y}_j$ are recorded in an ordered list\footnote{For $\ell = L + 1$, the ordered list $\mathcal{F}_j$ contains only numbers that correspond to the individual arrival times; but where $\ell = 1, \dots, L$, the list contains tuples.}, $\mathcal{F}_j$ (Algorithm \ref{__label__4f095ab3883a4ddfb2954e2fa11f65f9});
\item during the second phase, a sample path for process $\boldsymbol{Z}$ is generated using the MNRM. The waiting times for Poisson process $\mathcal{Y}_j$ are determined using ordered list $\mathcal{F}_j$ (Algorithm \ref{__label__bb3ee301d02b4d5687993b2472a47522}).
\end{itemize} 

\begin{algorithm}[!htb]
\caption{\protect\vphantom{$A^A$}Phase two of the CPM-coupled DM and tau-leap method.  This simulates an approximate sample path from the Poisson processes stored during phase one.\protect\vphantom{$A_A$}}  \label{__label__bb3ee301d02b4d5687993b2472a47522}
\begin{algorithmic}[1]
  \Require initial conditions, $\bs{Z}(0)$, terminal time, $T$, and ordered lists, $\mathcal{F}_j$ ($j = 1, \dots, M$).\protect\vphantom{$A_A^A$}
  \State set $\bs{Z} \leftarrow \bs{Z}(0)$, $t \leftarrow 0$, $t^* \leftarrow \tau$ 
  \State for each $R_j$, set $P_j \leftarrow 0$
  \State for each $R_j$, set $T_j$ to be the first element of list $\mathcal{F}_j$, then delete the first element of $\mathcal{F}_j$
  \State for each $R_j$, calculate propensity values $p_j(\bs{Z})$ 
  \Loop
  \State for each $R_j$, calculate $\Delta_j$ as \begin{equation} \Delta_j = \frac{T_j - P_j}{p_j}  \end{equation}
  \State set $\Delta \leftarrow \min_j \Delta_j$, and $k \leftarrow \text{argmin}_j \Delta_j$
  \If{$t + \Delta > T$}
  \State \textbf{break}
  \ElsIf {$t + \Delta > t^*$}
  \State set $t \leftarrow t^*$, and $t^* \leftarrow t^* + \tau$
  \State for each $R_j$, set $P_j \leftarrow P_j + p_j \cdot (t^* - t)$, then recalculate propensity $p_j$  
  \Else
  \State set $\bs{Z}(t + \Delta) \leftarrow \bs{Z}(t) + \bs{\nu}_k$, set $t \leftarrow t + \Delta$, and for each $R_j$, set $P_j \leftarrow P_j + p_j \cdot \Delta$
  \If {$\mathcal{F}_k \neq \emptyset$}
  \State let $u$ be the first element of $\mathcal{F}_k$: set $T_k \leftarrow T_k + u$, and delete the first element of $\mathcal{F}_k$
  \Else
  \State generate $u \sim \text{Exp}(1)$, then set $T_k \leftarrow T_k + u$ 
  \EndIf
  \EndIf
  \EndLoop
 \end{algorithmic}
\end{algorithm}

\section{Numerical experiments} \label{__label__439a9eb968c54826b8730e3976b4d497}
In this section, we consider {four} representative case studies. In each case, we will test our two tau-leap based multi-level coupling methods: the SPM and the CPM. We will also compare the SPM and CPM implementations of the multi-level method with the regular DM. 

Algorithmic performance will depend on the chosen algorithm parameters: $L$, $\mathcal{M}$ and $\tau_0$. A simple search procedure can easily discard clearly inefficient algorithm parameter choices, leaving us with a range of possible values for $L$, $\mathcal{M}$ and $\tau_0$ that should be investigated in greater detail. Further details are provided in reference \citep{__ref__b7cd69d78ffc4c2ba54f2f3e26b5d981}. By using the algorithm parameters earmarked for further investigation, we compare the computational performance of the SPM and CPM variance reduction methods on the multi-level method. We will therefore compare the relative performance of our algorithms using a wide range of choices of  $L$, $\mathcal{M}$ and $\tau_0$ that might be implemented by an end-user of the algorithm.

Our sample paths have been generated by using \texttt{C++}, according to the 2011 standard (\texttt{C++11}). Where possible, the \texttt{C++} Standard Template Library has been used, and all calculations have been performed with double precision. Sample paths were generated by using an \texttt{AMD FX-4350} CPU, with an advertised clock-speed of 4.2 GHz.

\subsection{Case study 1: a gene regulatory network model}
The first case study is a model of gene expression that has been used by \citet{__ref__5f82fa75a5314b0484ebeb52de9e2750} to test the multi-level method. In this model, mRNA is produced, and this leads to the production of proteins. The proteins bind to form dimers, and the mRNA and protein molecules may also decay. The biochemical reaction network is stated as:\begin{equation}
R_1: \, \emptyset\,\,\xrightarrow{25}\,\,M;
\quad
R_2: \, M\,\,\xrightarrow{1000}\,\,{ M+P};
\quad
R_3: \, P+P\,\,\xrightarrow{0.001}\,\,D; \label{__label__9ef7856d2dc0477b86427ec13f7ce37a}
\end{equation} \begin{equation*}
R_4: \, M\,\,\xrightarrow{0.1}\,\,\emptyset;
\quad
R_5: \, P\,\,\xrightarrow{1}\,\,\emptyset.
\end{equation*}  We write the numbers of mRNA, protein and dimer molecules at time $t$, respectively, as $\boldsymbol{X}(t)=[X_1(t),X_2(t),X_3(t)]^T$ and consequently the initial condition can be expressed as $\boldsymbol{X}(0)=[0,0,0]^T$. The parameters are as supplied, and are dimensionless. We will estimate the dimer population at terminal time $T=1$: this is given by $\mathbb{E}[X_3(1)]$. As a reference, this calculation can be undertaken with the DM. It takes a little over two hours ($7223$ seconds) to generate the $4.8\times10^6$ sample paths required for the estimate of $\mathbb{E}[X_3(1)] = 3714.3 \pm 1.0$. Here `$\pm$ 1.0' indicates a 95\% confidence interval of semi-length 1.0. 

We now use the multi-level method to estimate $\mathbb{E}[X_3(1)]$ to the same level of accuracy. We test both the SPM and CPM implementations of the multi-level method. In 

\begin{figure}[!htb]
\centering

\hrulefill
\vspace{5mm} 

\includegraphics[width=.8\textwidth]{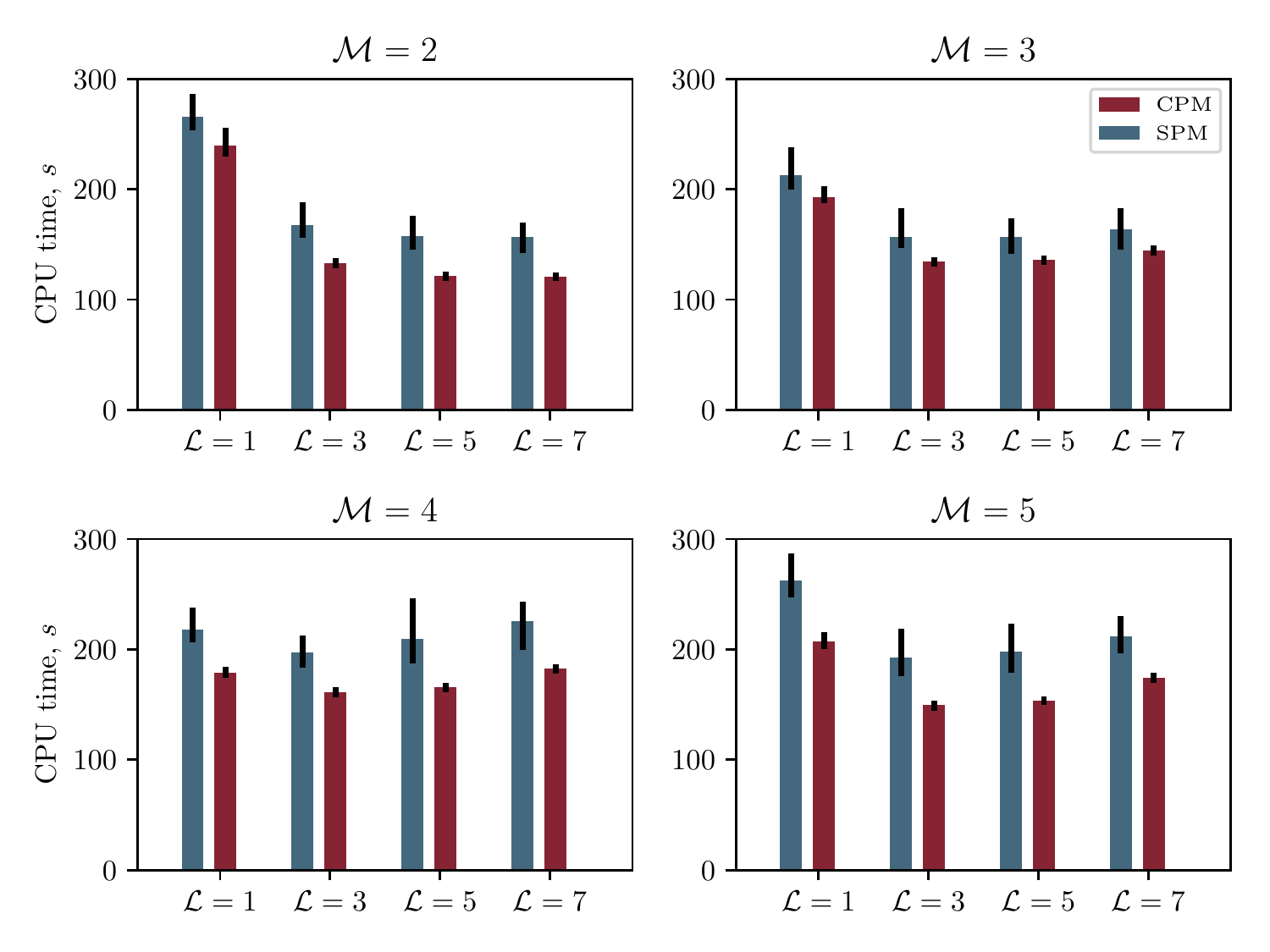}  

\caption{The average CPU time required by the multi-level method to estimate $\mathbb{E}[X_3(1)]$ for System \eqref{__label__9ef7856d2dc0477b86427ec13f7ce37a}. We vary $\mathcal{M}$ and $L$; the estimator is unbiased. The black bars indicate the range occupied by the $10$-th to $90$-th percentiles of the data. The values of $\tau_0$ are: $\mathcal{M} = 2 \Rightarrow \tau_0 = 1/8$; $\mathcal{M} = 3 \Rightarrow \tau_0 = 1/9$; $\mathcal{M} = 4 \Rightarrow \tau_0 = 1/16$ and $\mathcal{M} = 5 \Rightarrow \tau_0 = 1/5$.} \label{__label__c63d5f2168f94acc80acabbc9cf63231}
\hrulefill
\end{figure}

 \ref{__label__c63d5f2168f94acc80acabbc9cf63231}, we show the average CPU time taken by the respective multi-level method implementations. The values of  $L$, $\mathcal{M}$ and $\tau_0$ are varied; each average is computed over $100$ test runs of the complete multi-level algorithm. Our fastest CPM configuration takes, on average, $120.9$ seconds to estimate the dimer population (using $\mathcal{M} = 2$, $\tau_0 = 1/8$ and $L = 7$), and our most efficient SPM implementation takes, on average, $156.5$ seconds to estimate the same quantity (using $\mathcal{M} = 3$, $\tau_0 = 1/9$ and $L = 5$). For System \eqref{__label__9ef7856d2dc0477b86427ec13f7ce37a}, our most efficient SPM-implementation of the multi-level method therefore takes approximately 29\% longer to run than the comparable CPM implementation. In fact, for each test case that we considered, when compared with the CPM, the SPM implementation requires more CPU time to estimate the required summary statistic. 

Whilst the CPM implementation can reduce the average duration of multi-level Monte Carlo simulation, another key benefit is that its run-time is far more predictable than the run-time of the SPM implementation. The black lines in Figure \ref{__label__c63d5f2168f94acc80acabbc9cf63231} indicate the range occupied by the $10$-th to $90$-th percentiles of the total CPU time. It is clear that the CPU times of the CPM implementation are very tightly clustered around the mean CPU time, whilst, for the SPM, the CPU times display a higher variance. Moreover, the lower run-time of the CPM means that, even if a sub-optimal set of algorithm parameters is used, then the effect on the CPU time required is somewhat limited. 

\begin{figure}[!htb]
\centering

\hrulefill
\vspace{5mm} 

\includegraphics[width=.8\textwidth]{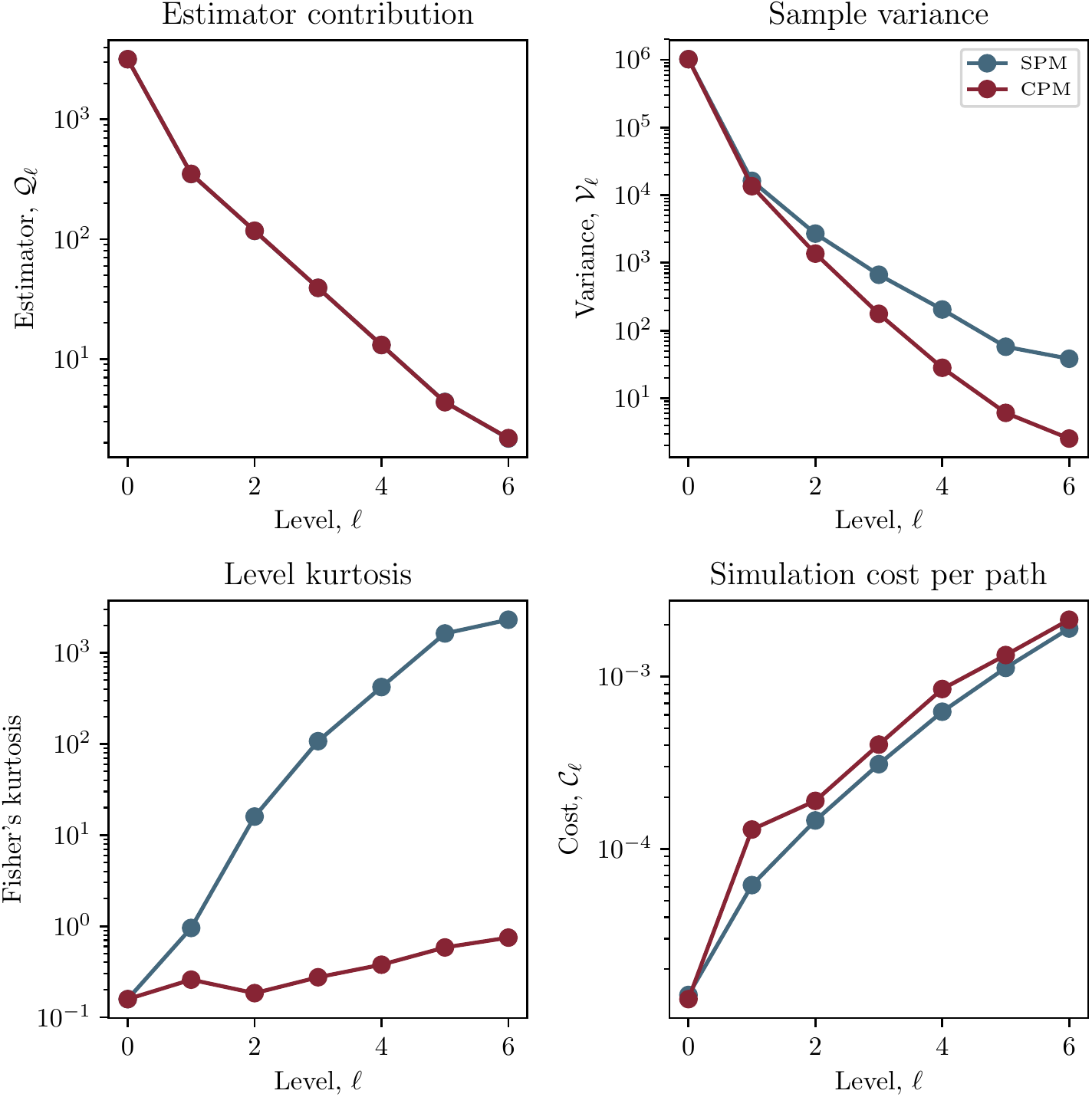}  
 \caption{The expected mean, sample variance, normalised kurtosis and CPU time for each level $\ell$ in our multi-level simulator for System \eqref{__label__9ef7856d2dc0477b86427ec13f7ce37a}. We have taken $\tau_0 = 1/9$, $\mathcal{M} = 3$ and $L = 5$, and used $10^5$ sample paths on each level. Note that the mean values for the SPM and CPM shown in the upper-left diagram overlap.} \label{__label__693cb36192a648619a84cfbbbdded4f9}
\hrulefill
\end{figure}

To compare the SPM with the CPM in more detail, we will need to concentrate on specific values of $\tau_0$, $\mathcal{M}$ and $L$. We will not hand-pick algorithm parameters that give the CPM any potential advantage over the SPM. We take $\tau_0 = 1/9$, $\mathcal{M} = 3$ and $L = 5$. With the aforementioned algorithm parameters, the CPM multi-level algorithm takes an average of $136.0$ seconds to run ($15.1$ seconds slower than the fastest CPM configuration), whilst the SPM method runs in an average of $156.5$ seconds (the fastest SPM configuration). In this case, the SPM takes approximately 15\% longer to run than the CPM method. 

In Figure \ref{__label__693cb36192a648619a84cfbbbdded4f9}, we show the empirical mean, sample variance, kurtosis and CPU time of each level estimator, $\mathcal{Q}_\ell$, when each of the SPM and CPM are used. The same base level ($\ell = 0$) is used for the SPM and CPM. The variances, $\mathcal{V}_\ell$, are substantially lower when the CPM, and not the SPM, is used. However, the CPU time taken to generate each sample, $\mathcal{C}_\ell$, is typically higher when the CPM is used. {For the higher levels, the use of the SPM results in a very high kurtosis (this is a consequence of a `catastrophic decoupling' which arises through the use of the SPM -- see reference \citep{__ref__b7cd69d78ffc4c2ba54f2f3e26b5d981} for an explanation).} A high kurtosis means that more of the sample variance can be attributed to infrequent but substantial deviations from the mean (as compared with frequent but modestly sized deviations from the mean). 

The effect of using the CPM implementation on the overall CPU time is outlined as follows. Subject to choosing $\mathcal{N}_\ell$ according to Equation \eqref{__label__b0d92eb1d30945e39fbbb5b435987319}, the total CPU time is given by Equation \eqref{__label__436d1497303048038259cf5f5c78ec9c}, i.e. \begin{equation*}
 \frac{1}{\varepsilon^2} \left\{\sum_{\ell=0}^{{L (+1)}} \sqrt{\mathcal{C}_\ell\cdot\mathcal{V}_\ell}\right\}^2.
\end{equation*} When the CPM is implemented, different values of $\mathcal{V}_\ell$ and $\mathcal{C}_\ell$ are inserted into Equation \eqref{__label__436d1497303048038259cf5f5c78ec9c}, and the result is that, with this case study, the total CPU time is reduced. 

We now explore the variation in CPU times between successive iterations of the multi-level method. Estimated values of $\mathcal{V}_\ell$ and $\mathcal{C}_\ell$ are used to populate Equation \eqref{__label__b0d92eb1d30945e39fbbb5b435987319}, and variations in these estimates cause variations in the CPU time expended by the multi-level method. There are two competing approaches for estimating $\mathcal{V}_\ell$ and $\mathcal{C}_\ell$: \begin{itemize}
\item the `one-step calibration' approach can be used. A small number of initial sample paths are generated for each level (for example, $10^2$ or $10^3$ sample paths), and then $\mathcal{V}_\ell$ and $\mathcal{C}_\ell$ are estimated. Equation \eqref{__label__b0d92eb1d30945e39fbbb5b435987319} is then evaluated, and the requisite number of sample paths is generated. This is the approach used by \citet{__ref__5f82fa75a5314b0484ebeb52de9e2750};
\item a repeated update, `dynamic calibration' approach can also be used. In this case, the estimated values of $\mathcal{V}_\ell$ are repeatedly updated as the algorithm progresses, and the required number of simulations on each level is refined. Where appropriate, additional sample paths can be generated. A detailed explanation is contained within our earlier work~\citep{__ref__b7cd69d78ffc4c2ba54f2f3e26b5d981}.

\end{itemize} The CPU times shown in Figure \ref{__label__c63d5f2168f94acc80acabbc9cf63231} were generated with the `dynamic calibration' procedure. We now evaluate the effect of using the CPM on the one-step and dynamic calibration procedures.

\begin{figure}[!b]
\centering

\hrulefill
\vspace{5mm} 

\includegraphics[width=.8\textwidth]{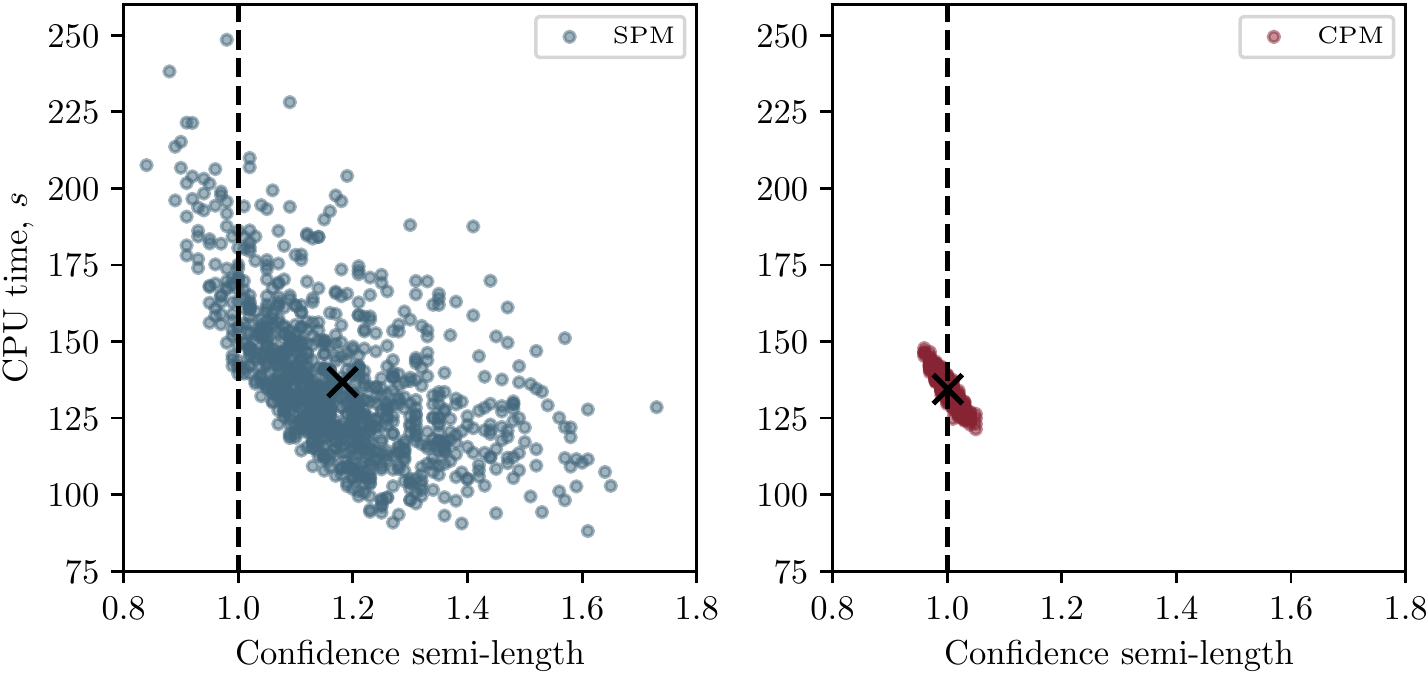}  
 \caption{The full multi-level method is run 1000 times to estimate $\mathbb{E}[X_3(1)]$ for System \eqref{__label__9ef7856d2dc0477b86427ec13f7ce37a} by following the `one-step calibration' approach, and using both the SPM and CPM. The CPU times are plotted against the semi-length of the confidence intervals attained. The black crosses represent the mean values of the data. The target confidence interval size is shown with a dashed line. The CPM is clearly superior.} \label{__label__88a595b0ad33412b8eefdae17432a97c}
\hrulefill
\end{figure}

\textbf{One-step calibration.} We first generate $\mathcal{N} = 10^2$ sample paths from which to estimate $\mathcal{V}_\ell$ and $\mathcal{C}_\ell$. In Figure \ref{__label__88a595b0ad33412b8eefdae17432a97c} we compare the effect of using the one-step calibration procedure with the SPM and the CPM. For each of the SPM and CPM, the entire multi-level method is run, from start until finish,  $1000$ times. We show the absolute CPU time, and plot this against the resultant confidence interval semi-length (as before, we aim for a confidence interval of semi-length $1.0$). 

Our results indicate that the CPM implementation is far more likely to achieve the required estimator variance than the SPM implementation, and to do so with a broadly comparable CPU time. The mean CPU times for the SPM and CPM are 136.7 and 134.4 seconds, respectively. Regrettably, the required confidence interval semi-length is not necessarily achieved; and it transpires that the estimated sample variances (based on $\mathcal{N} = 10^2$ initial paths) are not sufficiently accurate, and therefore too few sample paths are generated. The issue is far more pronounced for the SPM implementation as the required confidence interval is attained only 8\% of the time. With the CPM, the proportion of runs that achieve the required confidence intervals rises substantially to 58\%. 

\begin{figure}[!b]
\centering

\hrulefill
\vspace{5mm} 

\includegraphics[width=.8\textwidth]{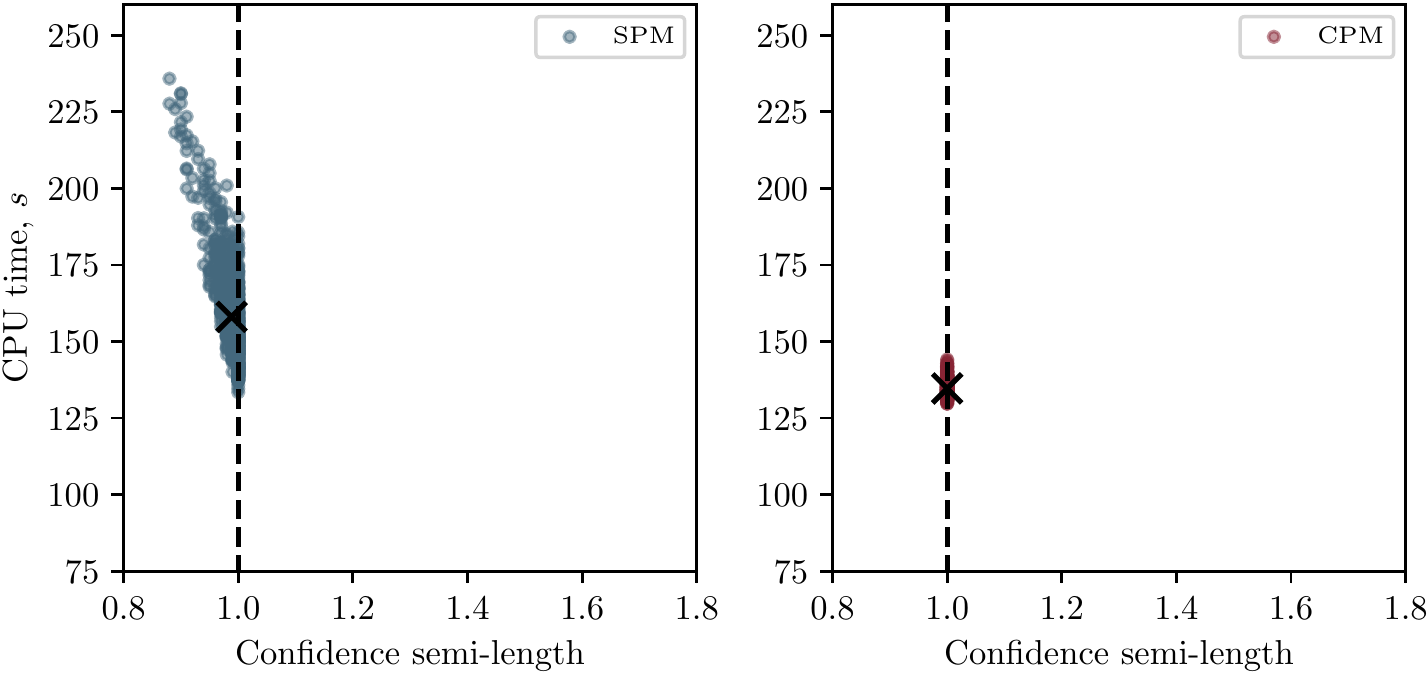}  
 \caption{The full multi-level method is run 1000 times to estimate $\mathbb{E}[X_3(1)]$ for System \eqref{__label__9ef7856d2dc0477b86427ec13f7ce37a} by following the `dynamic calibration' approach, and using both the SPM and CPM. The CPU times are plotted against the semi-length of the confidence intervals attained. The black crosses represent the mean values of the data. The target confidence interval size is shown with a dashed line. This approach is clearly preferable to that shown in Figure \ref{__label__88a595b0ad33412b8eefdae17432a97c}. The CPM still outperforms the SPM.} \label{__label__8ab3fce2ecab4f01aaa5278947b29da6}
\hrulefill
\end{figure}

\textbf{Dynamic calibration.} The difficulties encountered with estimated sample variances, $\mathcal{V}_\ell$, can be mitigated by using the `dynamic calibration' procedure: after the sample paths for each level are completed, the variance estimates can be updated as appropriate, and Equation \eqref{__label__b0d92eb1d30945e39fbbb5b435987319} used to recalculate the number of sample paths required for each level estimator. If the revised number of sample paths required for a given level is lower than the number of sample paths already generated, a `variance re-allocation' procedure can be followed, so that the sundry sample paths are not wasted. We follow the procedure outlined in our earlier work~\citep{__ref__b7cd69d78ffc4c2ba54f2f3e26b5d981} a total of 1000 times (we used $\mathcal{N} = 10^2$ initial paths to start the dynamic algorithm), we plot our results in Figure \ref{__label__8ab3fce2ecab4f01aaa5278947b29da6}. The average CPU times of the SPM and CPM are now 158.0 seconds and 134.7 seconds, respectively\footnote{The averages differ slightly from those presented in Figure \ref{__label__c63d5f2168f94acc80acabbc9cf63231}: the sample size is different.}. The average CPU time required by the CPM is broadly similar to the CPU time required without variance reallocation (see Figure \ref{__label__88a595b0ad33412b8eefdae17432a97c}), but the CPU time for the SPM method increases significantly.

Therefore, in this case, the CPM remains superior to the SPM. A CPM implementation of the multi-level method results in a decreased CPU time, when compared with the SPM. The CPM is reliable, and can be used with sample variances estimated using a small number of preliminary sample paths.

\subsection{Case study 2: Lotka-Volterra dynamics} \label{__label__5482aee58807473b9347296f4efb6347}
In this second case study, we evaluate the performance of the multi-level method with a stochastic analogue of the Lotka-Volterra system~\citep{__ref__39b5170e8a0949e99c877921832cdd3f}. The population dynamics of a predator, $A$, and its prey, $B$, are considered. The following reaction channels are defined: \begin{equation}
R_1: \, A\,\,\xrightarrow{10}\,\,\emptyset;
\quad
R_2: \, {A+B}\,\,\xrightarrow{0.01}\,\,2A;
\quad
R_3: \, B\,\,\xrightarrow{10}\,\,2B. \label{__label__f677e98bcf0242e29d719f07a0be9356}
\end{equation} Initially, the population of $A$, $X_1$, and the population of $B$, $X_2$, are both set to equal $1200$. We estimate the population levels of System \eqref{__label__f677e98bcf0242e29d719f07a0be9356} at time $T=3$.

System \eqref{__label__f677e98bcf0242e29d719f07a0be9356} clearly exhibits oscillatory dynamics, with the amplitude of the oscillations being highly unstable~\citep{__ref__77854e0e556e40b2bec0a7e868cb9542}. To our knowledge, the multi-level method has yet to be successfully applied to a system that exhibits such dynamics, even over a short time-interval. The average predator population, given by $\mathbb{E}[X_1(T)]$, can be estimated with the DM. This calculation takes a little over $25$ minutes (1523 seconds), and, with $1.94\times 10^5$ sample paths, we estimate $\mathbb{E}[X_1(T)] = 783.4 \pm 1.0$. 

\begin{figure}[ht]
\centering

\hrulefill
\vspace{5mm} 

\includegraphics[width=.6\textwidth]{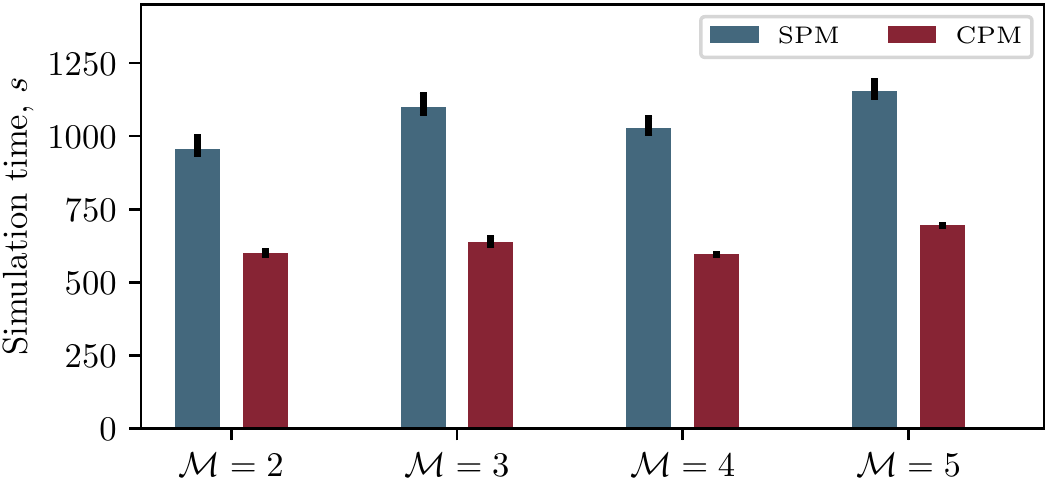}  
 \caption{The average CPU time required by the multi-level method to estimate $\mathbb{E}[X_1(30)]$ for System \eqref{__label__f677e98bcf0242e29d719f07a0be9356}. We vary $\mathcal{M}$ and $L$; the estimator is unbiased. The black bars indicate the range occupied by the $10$-th to $90$-th percentiles of the data. The values of $\tau_0$ are: $\mathcal{M} = 2 \Rightarrow \tau_0 = T\cdot2^{-12}$; $\mathcal{M} = 3 \Rightarrow \tau_0 = T\cdot3^{-8}$; $\mathcal{M} = 4 \Rightarrow \tau_0 = T\cdot2^{-12}$ and $\mathcal{M} = 5 \Rightarrow \tau_0 = T\cdot2^{-5}$. In each case, taking $L=1$ is optimal.}
  \label{__label__76aec3bc9cbd4abc8750892116147278}
\hrulefill
\end{figure}

We now use both the SPM- and CPM-controlled multi-level method to estimate  $\mathbb{E}[X_1(T)]$. In Figure \ref{__label__76aec3bc9cbd4abc8750892116147278} we show the average CPU times for a range of choices of the refinement factor, $\mathcal{M}$. In each case, we show the average CPU time achieved with the most efficient choice of $L$ and $\tau_0$ that we have found. The most efficient CPM algorithm we found requires $600.2$ seconds of CPU time; this is $37\%$ faster than the most efficient SPM algorithm, which requires $957.1$ seconds. For each choice of $\mathcal{M}$, the most efficient SPM algorithm took substantially longer to run than the most efficient CPM method. Moreover, if we compare the most efficient sets of algorithm parameters for the CPM and SPM, with the DM, then the CPM method is $2.5$ times faster than the DM, whereas the SPM by itself is 37\% faster than the DM. 

\subsection{Case study 3: a logistic growth model} \label{__label__7a0afc2c2d5c47ee8f44776d1bbdd121}
\begin{figure}[htb]
\centering

\hrulefill
\vspace{5mm} 

\includegraphics[width=.6\textwidth]{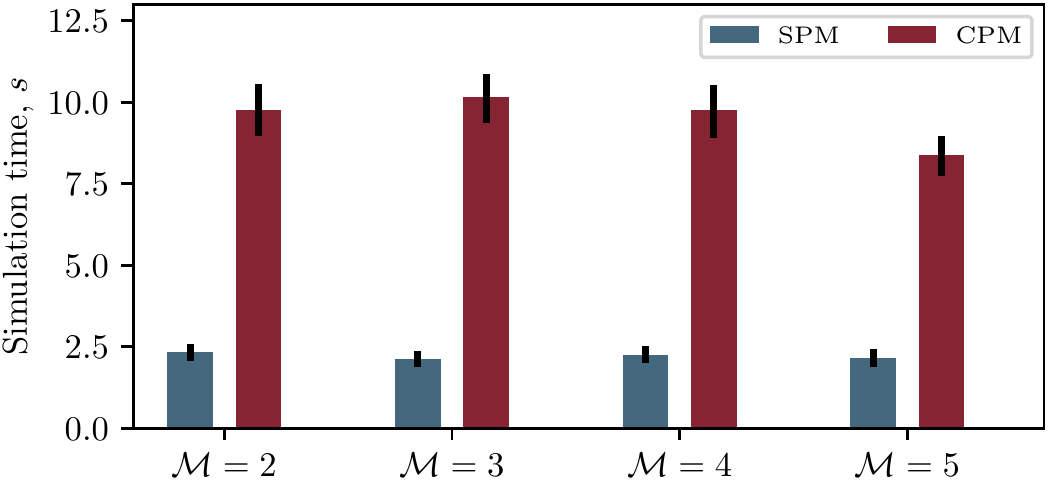}  
 \caption{The average CPU time required by the multi-level method to estimate $\mathbb{E}[X(3)]$ for System \eqref{__label__4115405a192647ac90ca3adb9f1ad68f}. The black bars indicate the range occupied by the $10$-th to $90$-th percentiles of the data. The values of $\tau_0$ and $L$ are individually optimised for each of the SPM and CPM.} \label{__label__88311f599e8e458cbdbcb44c0f4246e0}
\hrulefill
\end{figure}

The third case study is of a stochastic logistic growth model that comprises one species, and the following two reaction channels:\begin{equation}
R_1: \, A\,\,\xrightarrow{10}\,\,2A;
\quad
R_2: \, 2A\,\,\xrightarrow{0.01}\,\,{A}. \label{__label__4115405a192647ac90ca3adb9f1ad68f}
\end{equation} Initially, the population of $A$ is given as\footnote{As there is only one species, we will suppress the subscript and work with $X(t)$ instead of $X_1(t)$.} $X(0) = 50$. We will simulate System \eqref{__label__4115405a192647ac90ca3adb9f1ad68f} until a terminal time $T=3$, and estimate the mean population at that time, \begin{equation}
\mathcal{Q}_1 = \mathbb{E}[X(T)]. 
\end{equation}
The DM estimates $\mathcal{Q}_1 = 999.6 \pm 1.0$ using 3800 sample paths in $16.2$ seconds. We have run the multi-level algorithm with a wide variety of algorithm parameters (i.e. $L$, $\mathcal{M}$, and $\tau_0$) on this system. Our efforts are briefly summarised in Figure \ref{__label__88311f599e8e458cbdbcb44c0f4246e0}, which is constructed as follows: for each choice of $\mathcal{M}$, we find the optimal $L$ and $\tau_0$ for each of the SPM and CPM. 

We find that the multi-level method can be effectively implemented with the SPM, but the CPM approach is less efficient for this example. The SPM can estimate $\mathcal{Q}_1$ to a 95\% confidence interval of semi-length $1.0$ within $2.1$ seconds, which means that the DM requires 7.7 times as long to perform the calculation to the same level of accuracy. Our best result for the CPM is less efficient: our implementation requires $9.8$ seconds of CPU time to estimate $\mathcal{Q}_1$ to within the required statistical accuracy. Whilst the CPM substantially is slower than the SPM, the CPM is still faster than the DM. 

\begin{figure}[htb]
\centering

\hrulefill
\vspace{5mm} 

\includegraphics[width=.6\textwidth]{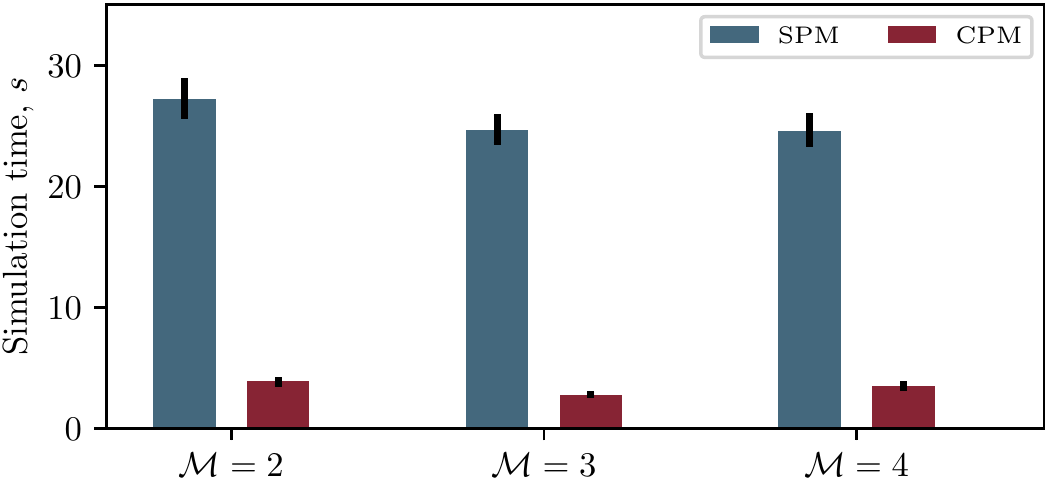}  
 \caption{The average CPU time required by the multi-level method to estimate $\mathcal{Q}_2 = \mathbb{E}\left[  \int_0^T X(t) \, \wrt{t} /T \, \right]$ for System \eqref{__label__4115405a192647ac90ca3adb9f1ad68f}. The black bars indicate the range occupied by the $10$-th to $90$-th percentiles of the data. The values of $\tau_0$ and $L$ are individually optimised for each of the SPM and CPM. } \label{__label__9b533c3b686c46f7abedc8563133b922}
\hrulefill
\end{figure}

{ For a second summary statistic, consider the time-averaged population, \begin{equation}
\mathcal{Q}_2 = \mathbb{E}\left[ \frac{1}{T}  \int_0^T X(t) \, \wrt{t} \, \right], \end{equation} for which the DM estimates $\mathcal{Q}_2 = 899.91 \pm 0.10$. This calculation requires $3.3\times10^4$ sample paths and takes $136.9$ seconds of CPU time to generate. 

To compare the DM against the SPM and CPM implementations of the multi-level algorithm, we evaluate the algorithm's performance with a wide variety of algorithm parameters. Our efforts are briefly summarised in Figure \ref{__label__9b533c3b686c46f7abedc8563133b922}, which is constructed as before: for each choice of $\mathcal{M}$, we find the optimal $L$ and $\tau_0$ for each of the SPM and CPM. 

We find that the multi-level method can be effectively implemented with the SPM, but, for this second summary statistic, $\mathcal{Q}_2$, the CPM approach is even more efficient. The SPM can estimate $\mathcal{Q}_2$ to a 95\% confidence interval of semi-length $1.0$ within $24.7$ seconds, which means that the DM requires $5.5$ times as long to perform the calculation, to the same level of accuracy. Our best result for the CPM requires only $2.8$ seconds of CPU time to estimate $\mathcal{Q}_2$ to within the required statistical accuracy. Thus, the CPM is $8.8$ times more efficient than the SPM, and the DM requires approximately $49.2$ times as long to perform the same calculation. }

{
\subsection{Case study 4}
We now consider a final case study. The MAPK pathway is a chain of proteins in the cell that communicates a signal from a receptor on the cell surface to the DNA in the nucleus. This case study is of a model of the MAPK cascade, which is involved in a variety of signalling processes that govern transitions relating to the phenotype of a cell~\citep{__ref__6c309c4c12884051a38b7be20b047e02}. 

The model of \citet{__ref__cfcfd0d4f21146a3bb4809687529f7702} represents a MAPK cascade, and comprises ten coupled Michaelis-Menten schemes, with $N=22$ species and $M=30$ reactions (see Figure \ref{__label__ac22288b04544bb3a22d937442196267}. A Michaelis-Menten scheme is constructed as follows: there are four species and three reaction channels within the scheme~\citep{__ref__6c309c4c12884051a38b7be20b047e02}. The species are substrate (`S'), enzyme (`E'), complex (`ES') and product (`P'). The reaction channels are as follows: \begin{equation}R_1: \, E + S \,\,\autorightarrow{\footnotesize{$r_1$}{}} \,\, ES;                                                                                   
\quad
R_2: \, ES \,\,\autorightarrow{\footnotesize{$r_{-1}$}}{}\,\, E + S;
\quad
R_3: \, ES \,\,\autorightarrow{\footnotesize{$r_{2}$}}{}\,\, E + P. \label{__label__57da3f9daeca4a799ef1c83ccacd853c}
\end{equation} A quasi-steady state assumption is applied to reduce the computational complexity associated with simulating the reaction network. This reduces the scheme to two species: substrate (`S') and product (`P'). The three reaction channels described by System \eqref{__label__57da3f9daeca4a799ef1c83ccacd853c} are reduced into a single reaction channel, which is given as  \begin{equation}
R_*: \, S \,\,\autorightarrow{\footnotesize{$k(S)$}}{}\,\, P, \label{__label__15617f2083d24d18950912e54d91db48}
\end{equation}  where the reaction rate follows Michaelis-Menten kinetics,
\begin{equation}\label{__label__b4604c2f2a8441949ad1f8c55c813b85}
k(S) = \frac{k_2 E_0}{S + \frac{r_{-1} + r_2}{r_1}},
\end{equation} where $E_0$ represents the initial enzyme population. The substrate, enzyme and product molecules for each channel are as shown in Figure \ref{__label__ac22288b04544bb3a22d937442196267}. The reaction channels are therefore:\begin{equation*}
R_1: \, KKK \,\,\autorightarrow{\footnotesize{$k_1$}}{}\,\, KKK\textrm{-}P;
\quad
R_2: \, {KKK\textrm{-}P} \,\,\autorightarrow{\footnotesize{$k_2$}}{}\,\,{KKK};
\end{equation*}\vspace{-5mm}\begin{equation*}
R_3: \, KK\,\,\autorightarrow{\footnotesize{$k_3$}}{}\,\,KK\textrm{-}P;
\quad
R_4: \, {KK\textrm{-}P} \,\,\autorightarrow{\footnotesize{$k_4$}}{}\,\,{KK}; 
\end{equation*}\vspace{-5mm}\begin{equation}
R_5: \, KK\textrm{-}P\,\,\autorightarrow{\footnotesize{$k_5$}}{}\,\,KK\textrm{-}PP;
\quad
R_6: \, KK\textrm{-}PP \,\,\autorightarrow{\footnotesize{$k_6$}}{}\,\,{KK\textrm{-}P};
\label{__label__a6fd7a3c60894b8ea93a4437e69101cb} 
\end{equation}\vspace{-5mm}\begin{equation*}
R_7: \, K\,\,\autorightarrow{\footnotesize{$k_7$}}{}\,\,K\textrm{-}P;
\quad
R_8: \, K\textrm{-}P\,\,\autorightarrow{\footnotesize{$k_8$}}{}\,\,K;
\end{equation*}\vspace{-5mm}\begin{equation*}
R_9: \, K\textrm{-}P\,\,\autorightarrow{\footnotesize{$k_9$}}{}\,\,K\textrm{-}PP;
\quad
R_{10}: \, K\textrm{-}PP \,\,\autorightarrow{\footnotesize{$k_{10}$}}{}\,\,{K\textrm{-}P},
\end{equation*} where the $k_j$ are rate functions specified by Equation~\eqref{__label__b4604c2f2a8441949ad1f8c55c813b85}.

\begin{figure}[bth]
\centering

\hrulefill

\vspace{3mm}

\includegraphics[width=.6\linewidth]{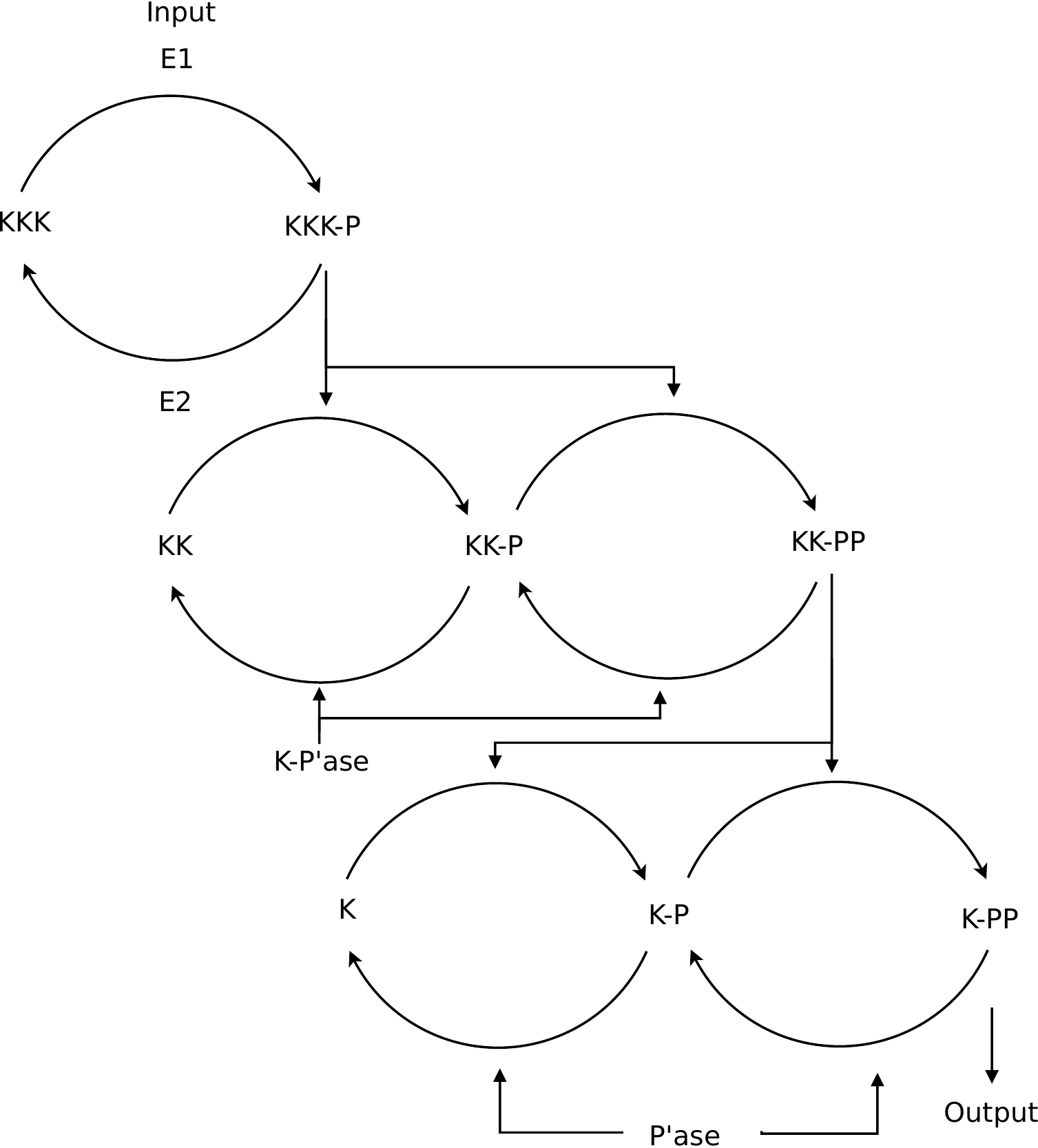}
\caption{A diagrammatic representation of the MAPK cascade. The text refers to chemical species; whilst the curved arrows represent Michaelis-Menten schemes. The arrow points from the substrate towards the product; the species on top of the arc indicates the enzyme. This diagram has been adapted from \citet{__ref__cfcfd0d4f21146a3bb4809687529f7702}.}
\label{__label__ac22288b04544bb3a22d937442196267}

\hrulefill

\end{figure}

We estimate the mean MAPK population (indicated by `K-PP' in System~\eqref{__label__a6fd7a3c60894b8ea93a4437e69101cb} and Figure~\ref{__label__ac22288b04544bb3a22d937442196267}) at a terminal time $T$. The initial conditions are detailed in Table \ref{__label__fc41108fb8f34967b0f86905e30d3d03}, and we take $T = 250$. Each Michaelis-Menten reaction is of the form $R_j: \,X \,\,\autorightarrow{\footnotesize{$k_j$}}{}\,\,{Y}$, and the function $k_j$ is expressed as $k_j = \alpha_j\cdot X / (X + \beta_j)$. For each reaction $R_j$, the initial enzyme populations give $\alpha_j$ and $\beta_j$ their values. The values that we use for $\alpha_j$ and $\beta_j$ (for $j = 1, \dots, 10$) are stated in Table \ref{__label__a98dc1b20cb4444ea674967abb3eb1fa}.

\begin{table}[bth]

\hrulefill

\vspace{3mm}

\centering
\begin{tabular}{|cc|cc|}
\hline \hline
\textbf{Species}	&	\textbf{Initial value}	&	\textbf{Species}	&	\textbf{Initial value} \\ \hline
\rule{0pt}{2ex}
$KKK$\rule{0mm}{4.25mm}	&$90$&	$KKK\textrm{-}P$	&	$10$ \\
$KK$	&$280$&	$KK\textrm{-}P$	&	$10$\\
$KK\textrm{-}PP$	&$10$&	$K$	&	$280$\\
$K\textrm{-}P$	&$10$&	$K\textrm{-}PP$	&	$10$ \\ \hline \hline
%\hline \hline      
\end{tabular}
\caption{The initial values for the MAPK cascade model detailed in \eqref{__label__a6fd7a3c60894b8ea93a4437e69101cb}.}
\label{__label__fc41108fb8f34967b0f86905e30d3d03}

%\hrulefill

\end{table} \begin{table}[tbh]
%\hrulefill

%\vspace{3mm}

\centering
\begin{tabular}{|cl|cl|}
\hline \hline
\textbf{Reaction}	&	\multicolumn{1}{c|}{\textbf{Parameters}} 	&	\textbf{Reaction}	&	\multicolumn{1}{c|}{\textbf{Parameters}} \\ \hline
\rule{0pt}{2ex}
$R_1$\rule{0mm}{4.25mm} & $\alpha_1 = 2.5$, $\beta_1 = 10$ & $R_2$ & $\alpha_2 = 0.25$, $\beta_2 = 8$ \\ 
$R_3$ & $\alpha_3 = 0.025$, $\beta_3 = 15$ & $R_4$ & $\alpha_4 = 0.75$, $\beta_4 = 15$ \\
$R_5$ & $\alpha_5 = 0.025$, $\beta_5 = 10$ & $R_6$ & $\alpha_6 = 0.75$, $\beta_6 = 15$ \\ 
$R_7$ & $\alpha_7 = 0.025$, $\beta_7 = 10$ & $R_8$ & $\alpha_8 = 0.5$, $\beta_8 = 15$ \\ 
$R_9$ & $\alpha_9 = 0.025$, $\beta_9 = 10$ & $R_{10}$ & $\alpha_{10} = 0.5$, $\beta_{10} = 15$ \\ \hline \hline
\end{tabular}
\caption{The parameters for the MAPK cascade model \eqref{__label__a6fd7a3c60894b8ea93a4437e69101cb}.}
\label{__label__a98dc1b20cb4444ea674967abb3eb1fa}
\hrulefill
\end{table}

If the DM is used, it takes approximately $64.8$ seconds to estimate the mean MAPK population at time $T = 250$ as $2683.16 \pm 0.99$. To study the performance of the multi-level method, we again investigate a range of algorithm parameters. Settling on the efficient choice of $\mathcal{M} = 4$, $\tau_0 = 1/16$ and $L = 3$, and running the multi-level a total of $100$ times for each case, we see that the SPM estimates the MAPK population within an average of $13.9$ seconds  (with the $10$-th percentile corresponding to 13.3 seconds, and the $90$-th percentile, 14.5 seconds), whilst the CPM takes an average of $12.1$ seconds (with the $10$-th percentile corresponding to 11.5 seconds, and the $90$-th percentile, 12.5 seconds). Whilst both the SPM and CPM are substantially more efficient than the DM, the CPM is approximately 14.9\% faster than the SPM. This demonstrates that, even with a relatively complicated reaction network, a significant reduction in simulation time can be achieved with the multi-level method. }

\section{Multi-level Monte Carlo with the R-leap method} \label{__label__4f2afd7baa2b4732a8ee837f2fc214bf}
In this section, we present a new implementation of the multi-level method. With a view to improving computational performance, we will use the R-leap method to construct an efficient multi-level algorithm. Once we have described our method, we demonstrate the performance of our algorithm with an example.

\subsection{Variance reduction with the R-leap method}
Our description of the R-leap multi-level method will mimic the description of the tau-leap multi-level method described in Section \ref{__label__471d7eaf487749a0885bc893309e52bd}. We write the summary statistic of interest, $\mathcal{Q}$, as the following telescoping sum: \begin{equation*}
\mathcal{Q} =\sum_{\ell=0}^{L + 1} \mathcal{Q}_\ell.
\end{equation*} The values of $\mathcal{Q}_\ell$, for $\ell = 0, \dots, L + 1$, are now determined as follows: \begin{itemize}
\item on the \emph{base level}, where $\ell = 0$, we generate sample paths using the R-leap method, with $\mathcal{K}_0$ reactions simulated at each step. We use a large value of $\mathcal{K}_0$, so that we can quickly produce the sample paths required to estimate $\mathcal{Q}_0$. The $r$-th such sample paths is labelled as $\bs{Z}_0^{(r)}$, and we use the scalar ${Z}_0^{(r)}$ to represent the point statistic, ${Z}_0^{(r)} \coloneqq f\big(\bs{Z}_0^{(r)}\big)$. Accordingly, the estimator for level $0$ is given by \begin{equation*}
 \mathcal{Q}_0 \coloneqq \mathbb{E}\Big[Z_0^{(r)}\Big] \approx \frac{1}{\mathcal{N}_0} \sum_{r = 1}^{\mathcal{N}_0}  Z_0^{(r)} ;\end{equation*}
\item the \emph{correction levels} ($\ell =1, \dots, L$) require the generation of pairs of sample paths, $[\bs{Z}_{{\ell-1}}$, $\bs{Z}_{\ell}]$. The `fine' sample path $\bs{Z}_{\ell}$ is generated using the R-leap method, where  $\mathcal{K}_\ell$ reactions are fired during each step. The `coarse' sample path $\bs{Z}_{{\ell-1}}$ is also generated with the R-leap method, but with  $\mathcal{K}_{\ell - 1}$ reactions at each step. As with the tau-leap multi-level method, we choose a refinement factor, $\mathcal{M}$, so that $\mathcal{K}_\ell = \mathcal{K}_{\ell-1} / \mathcal{M}$. The estimator for level $\ell$ is \begin{equation*}
 \mathcal{Q}_\ell \coloneqq \mathbb{E}\Big[Z_\ell^{(r)} - Z_{\ell-1}^{(r)}\Big] \approx \frac{1}{\mathcal{N}_\ell} \sum_{r = 1}^{\mathcal{N}_\ell} \left[ Z_\ell^{(r)} - Z_{\ell-1}^{(r)} \right];
\end{equation*}
\item finally, and optionally, the \emph{final correction level}, $L + 1$, removes all remaining bias:
\begin{equation*}
 \mathcal{Q}_{L + 1} \coloneqq \mathbb{E}\Big[X^{(r)} - Z_{L}^{(r)}\Big] \approx \frac{1}{\mathcal{N}_{L + 1}} \sum_{r = 1}^{\mathcal{N}_{L + 1}} \left[ X^{(r)} - Z_{L}^{(r)} \right].
\end{equation*}
\end{itemize} The sample paths for $\mathcal{Q}_0$ are performed with the regular R-leap method; pseudo-code is provided in Algorithm \ref{__label__a399905995f94d32a3f5d4befd66bb4d}. As before, we will use an algorithm that \emph{couples} sample paths, so that $\mathcal{Q}_1, \dots, \mathcal{Q}_{L+1}$ can be estimated with a low variance. Note that, unlike the tau-leap method implementation, a special algorithm for the `final estimator' (i.e. $\mathcal{Q}_{L+1}$) is not required. If we set $\mathcal{K}_{L+1} = 1$, then $\mathcal{Q}_{L+1}$ can be estimated with exactly the same method as $\mathcal{Q}_1, \mathcal{Q}_2, \dots, \mathcal{Q}_L$. As before, the multi-level method will only reduce computational costs if we can efficiently estimate $\mathcal{Q}_1, \dots, \mathcal{Q}_{L+1}$: we now discuss techniques for doing so.

When we implement the R-leap method within the multi-level scheme, we use a coupling method that combines elements of the CPM and the SPM. We start by referring to Section \ref{__label__fe56d15596ea45beb4d02983da69e032}, which states that at each step of the R-leap algorithm, two quantities are stochastically generated: \begin{enumerate}
\item the time-period covered by that step; 
\item the precise combination of reactions that fire during that time-period.

\end{enumerate} We will let sample paths $\bs{Z}_{\ell}$ (which we will call the fine path) and $\bs{Z}_{{\ell-1}}$ (which we will call the coarse path) advance by different time-periods at each step (point 1 above). A variance reduction technique is used to choose the time-periods that each sample path traverses. Then, to achieve maximal variance reduction, we will also ensure that, as far as possible, the same reactions fire in each sample path (point 2 above).

In order to simultaneously generate a pair of sample paths, $\big[\bs{Z}_{\ell}, \bs{Z}_{{\ell-1}}\big]$, at each step of the coupled simulation algorithm a total of $\mathcal{K}_{\ell}$ $( = \min\{\mathcal{K}_{\ell}, \mathcal{K}_{\ell-1}\}$) reaction events will take place in \emph{each} sample path. In particular: \begin{itemize} 
\item a Gamma variate, $\Delta = \Gamma(\mathcal{K}_\ell, 1)$, is generated. The time-period spanned by this step in each of the coarse and fine sample paths is then determined by a rescaling argument. In distribution, $\Gamma(\mathcal{K}, \theta) \sim \theta \cdot \Gamma(\mathcal{K}, 1)$. Therefore, the time-period for fine path is given by $\Delta / \sum_{j=1}^M p_j^F$, and for the coarse system by $\Delta / \sum_{j=1}^M p_j^C$; 
\item the precise combination of reactions is chosen as follows. For each of the $\mathcal{K}_\ell$ reactions that take place in \emph{each} sample path, the probability that it is a $R_j$ reaction is given by $p_j^C / \sum_{j'=1}^M p_{j'}^C$ (for the coarse path), and $p_j^F / \sum_{j'=1}^M p_{j'}^F$ (for the fine path). Our coupling method must therefore fire reactions with these probabilities. For each $R_j$ (for $j=1, \dots, M$), we define the probabilities $b_j^m$, where $m \in \{1, 2, 3\}$ as:
\begin{equation}
 b_j^1 = \min \left\{ \frac{p_j^C}{\sum_{j'=1}^M p_{j'}^C}, \frac{p_j^F}{\sum_{j'=1}^M p_{j'}^F} \right\}; \label{__label__458ab95eeeac4b7dbba9920ef4b5682f}
\end{equation}\begin{equation*}
                b_j^2 = \frac{p_j^C}{\sum_{j'=1}^M p_{j'}^C} - b_j^1;  \hspace{8mm}  b_j^3 =\frac{p_j^F}{\sum_{j'=1}^M p_{j'}^F} - b_j^1.
              \end{equation*}  We interpret each probability, $b_j^m$, where $m \in \{1,2,3\}$ by noting\footnote{Note that, for each $j$, at least one of  $b_j^2$ and $b_j^3$ will be zero.}: \begin{itemize}
              \item $b_j^1$ represents the probability reaction $R_j$ takes place in both the coarse and the fine paths;
              \item $b_j^2$ represents the probability reaction $R_j$ takes place in only the coarse  path;
              \item $b_j^3$ represents the probability reaction $R_j$ takes place in only the fine path;
              \end{itemize}
              
              As with the regular R-leap method, a conditional binomial method is used to choose the precise combination of reactions that take place. A total of $\mathcal{K}_\ell$ reaction events must take place in each of the coarse and the fine paths. We would, if possible, like the same reactions to fire in each sample path. Thus, we first consider the `common' reactions -- i.e. those that occur with probability $b_j^1$ (for $j=1, \dots, M$). Once the `common' reactions have been determined, then we will determine the reactions specific to either the coarse or the fine path. Note that, as we are coupling sample paths, we expect that the propensities of the fine and coarse paths are similar for each $R_j$, i.e. $p_j^F \sim p_j^C$, so that $p_j^F / \sum_{j'=1}^M p_{j'}^F \sim p_j^C / \sum_{j'=1}^M p_{j'}^C$. Following Equation \eqref{__label__458ab95eeeac4b7dbba9920ef4b5682f}, we conclude that $b^1_j \gg b^2_j, b^3_j$. Therefore, most events will be in the form of a `common' reaction, and so will take place in both the coarse and the fine sample paths.

\item the propensity values are updated as appropriate. The propensities associated with the fine path are updated at every step of the algorithm (i.e. after the required $\mathcal{K}_{\ell}$ reaction events have taken place in the fine path), whilst the propensities of the coarse path are updated every $\mathcal{M}$ steps of the algorithm (i.e. after $\mathcal{M}\cdot \mathcal{K}_\ell = \mathcal{K}_{\ell-1}$ reaction events have taken place in the coarse path). 
              \end{itemize}

As mentioned, once the common reactions have been completed, the remaining reactions events must be performed. The R-leap coupling method is presented as pseudo-code in Algorithm \ref{__label__8971cfd3502d481b87b09c06027af196}. We now proceed to present numerical results in Section \ref{__label__88f6fc6faf8045ddb126a0a8bf934098}. The computational performance of our new method is compared with the tau-leap multi-level method and traditional simulation methods.

\begin{algorithm}[!hb]
\caption{The coupled R-leap method. This simulates a pair of sample paths.\protect\vphantom{$A_A^A$}}
\label{__label__8971cfd3502d481b87b09c06027af196}
 \begin{algorithmic}[1]
  \Require initial conditions, $\bs{Z}(0)$, $\mathcal{K}$ ($=\mathcal{K}_\ell$), $\mathcal{M}$ ($=\mathcal{K}_{\ell-1} / \mathcal{K}_\ell$), and terminal time, $T$.\protect\vphantom{$A_A^A$}
  \State set $\bs{Z}^C \leftarrow \bs{Z}(0)$, $\bs{Z}^F \leftarrow \bs{Z}(0)$, $t^C \leftarrow 0$ and $t^F \leftarrow 0$
  \State for each $R_j$, calculate propensity values $p^C_j\big(\bs{Z}^C\big)$ and $p^F_j\big(\bs{Z}^F\big)$
  \State set flag $\leftarrow \texttt{false}$, iter $\leftarrow 0$
  \While {flag \textbf{is} \texttt{false}}
  \State set iter $\leftarrow$ iter $+$ $1$
  \State generate $\Delta \sim \Gamma(\mathcal{K}, 1)$
  \State set $t^C\leftarrow t^C + \Delta / \sum_{j=1}^M p_j^C$ and  $t^F\leftarrow t^F + \Delta / \sum_{j=1}^M p_j^F$
  \If {$\max\{t^C, t^F\} > T$}
  \State set $\mathcal{K} \sim \mathcal{B}\big(\mathcal{K}-1, \min\big\{\big(T - t^C\big)/\big(\Delta / \sum_{j=1}^M p_j^C\big), \big(T - t^F\big)/\big(\Delta / \sum_{j=1}^M p_j^F\big)-1\big)\big\}$
  \State set $\text{max}\{t^C, t^F\} \leftarrow T$
  \State set flag $\leftarrow \texttt{true}$
  \EndIf
  \State set $K^* \leftarrow \mathcal{K}$
  \State for each $R_j$, calculate probabilities $b^1_j$, $b^2_j$ and $b^3_j$ according to Equations \eqref{__label__458ab95eeeac4b7dbba9920ef4b5682f}
  \For {$j = 1, \dots, M$}
  \State generate $K_{j1} \sim \mathcal{B}\big(K^*, b^1_j/\big(1-\sum_{j'=0}^{j-1} b^1_{j'}\big) \big)$ and set $K^* \leftarrow K^* - K_j$
  \EndFor
  \State set $K^{*2} \leftarrow K^*$ and  $K^{*3} \leftarrow K^*$
  \For {$j = 1, \dots, M$}
  \State generate $K_{j2} \sim \mathcal{B}\big(K^{*2}, b^2_j/\sum_{j'=j}^Mb^2_{j'}\big)$ and set $K^{*2} \leftarrow K^{*2} - K_{j2}$
  \State generate $K_{j3} \sim \mathcal{B}\big(K^{*3}, b^3_j/\sum_{j'=j}^Mb^3_{j'}\big)$ and set $K^{*3} \leftarrow K^{*3} - K_{j3}$
  \EndFor 
  
  \State set $\bs{Z}^C \leftarrow \bs{Z}^C + \sum_{j=1}^M \big(K_{j1}+K_{j2}\big) \cdot \bs{\nu}_j$
  \State set $\bs{Z}^F \leftarrow \bs{Z}^F + \sum_{j=1}^M \big(K_{j1}+K_{j3}\big) \cdot \bs{\nu}_j$
  \State for each $R_j$, calculate propensity values $p^F_j\big(\bs{Z}^F\big)$ 
  \State \textbf{if} $\mathcal{M}$ divides iter, for each $R_j$, calculate propensity values $p^C_j\big(\bs{Z}^C\big)$ 
   \EndWhile
  \State \textbf{if} $t^C < T$  \textbf{then} use Algorithm \ref{__label__a399905995f94d32a3f5d4befd66bb4d} to simulate $\bs{Z}^C$ until time $T$. \textbf{end if}. 
   \State \textbf{if} $t^F < T$  \textbf{then} use Algorithm \ref{__label__a399905995f94d32a3f5d4befd66bb4d} to simulate $\bs{Z}^F$ until time $T$. \textbf{end if}. 

  \end{algorithmic}
\end{algorithm}

\subsection{Numerical experimentation} \label{__label__88f6fc6faf8045ddb126a0a8bf934098}
\begin{figure}[htb]
\centering

\hrulefill
\vspace{5mm} 

\includegraphics[width=.6\textwidth]{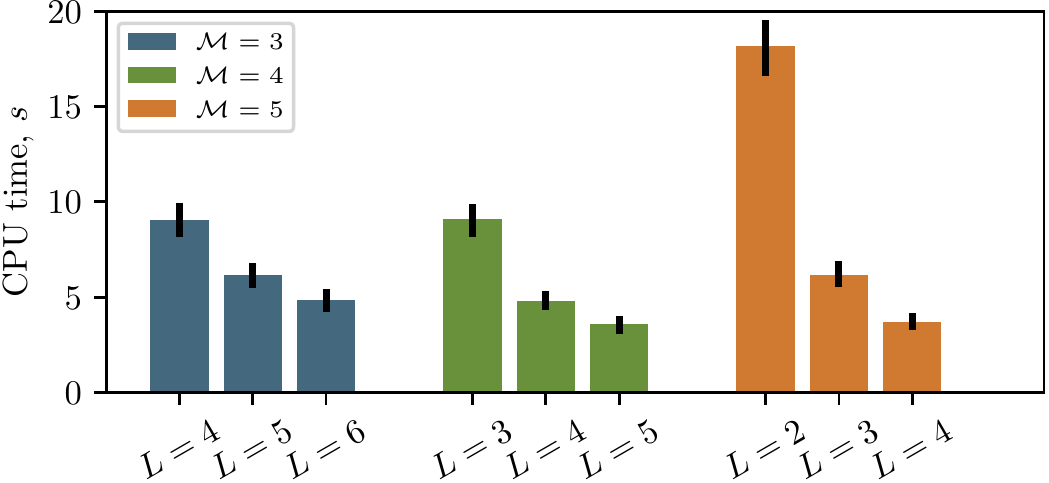}  
 \caption{The average CPU time required by the R-leap multi-level method to estimate $\mathbb{E}[X(3)]$ for System \eqref{__label__4115405a192647ac90ca3adb9f1ad68f}. We vary $\mathcal{M}$ and $L$; the estimator is unbiased. The black bars indicate the range occupied by the $10$-th to $90$-th percentiles of the data.} \label{__label__3232c2f409ef4b08a7171bd042d2a399}
\hrulefill
\end{figure}

We return to Case study 3 presented in Section \ref{__label__7a0afc2c2d5c47ee8f44776d1bbdd121}. The reaction channels are \begin{equation*}
R_1: \, A\,\,\xrightarrow{10}\,\,2A,
\quad
R_2: \, 2A\,\,\xrightarrow{0.01}\,\,{A},
\end{equation*} with initial conditions as described in Section \ref{__label__7a0afc2c2d5c47ee8f44776d1bbdd121}. We return to estimating $\mathcal{Q}_1 = \mathbb{E}[X(3)]$; for reference, we note that the DM estimates $\mathcal{Q}_1$ to be $999.6 \pm 1.0$ within 16.2 seconds. The R-leap multi-level method has been implemented with a range of choices of $\mathcal{M}$ (the refinement factor) and $L$ (that controls the number of levels). We seek an unbiased estimate, which means that $\mathcal{K}_0 = \mathcal{M}^{L+1}$, $\mathcal{K}_1 = \mathcal{M}^{L}$, $\dots$,  $\mathcal{K}_{L} = \mathcal{M}$, and $\mathcal{K}_{L+1} = 1$. Our results are collated in Figure \ref{__label__3232c2f409ef4b08a7171bd042d2a399}. The most efficient R-leap multi-level implementation estimates $\mathcal{Q}_1$ using $3.6$ seconds of CPU time. The DM method therefore takes approximately $4.5$ times longer than our R-leap multi-level method. 

Our results demonstrate that the R-leap multi-level method is a feasible alternative to the tau-leap multi-level method. If we consider the numerical performance of Case study 3 when the tau-leap multi-level method is used, it is clear that for this example, R-leap is superior to a CPM tau-leap method (which takes $9.8$ seconds to estimate $\mathcal{Q}$), but inferior to a SPM implementation (which we recall takes $2.1$ seconds to estimate $\mathcal{Q}$). The optimal method will depend on the particular reaction network and summary statistics of interest.

\section{Discussion} \label{__label__081a6f79524644a9b93459bfc2788cf2}
At the outset, we presented two related aims for this manuscript: firstly, to improve the robustness of the multi-level method; and, secondly, to improve the computational performance of the multi-level method. Through the use of the SPM and the CPM approaches, the multi-level method has the potential to dramatically reduce the CPU time required to carry out Monte Carlo simulation. Whilst the CPM method does not always outperform the SPM method (although it often does), we have demonstrated its effectiveness and reliability. The R-leap multi-level method has been demonstrated as a new and effective simulation technique. The computational performance of the multi-level method may still depend on the particular reaction network and the chosen summary statistic, but our new methods provide additional tools for accelerating stochastic simulation. 

In Case study 1, we demonstrated that, for the gene regulatory reaction network, the CPM is nearly 30\% more efficient than the SPM. Furthermore, we noted that the kurtoses of levels $\ell = 1, \dots, L + 1$ are substantially higher when the SPM is used, when compared with the CPM. {The high kurtoses make it difficult to estimate the sample variances of the SPM method.} There are two possible consequences: \begin{itemize}
\item the sample variance, $\mathcal{V}_\ell$, is an under-estimate. The effect is that the required confidence interval semi-length is not faithfully attained;
\item the sample variance $\mathcal{V}_\ell$, is an over-estimate. In this case, too many sample paths are generated, and the algorithm takes substantially longer to run.
\end{itemize} A more robust approach is therefore provided by the CPM. Over a wide range of algorithm parameters, the CPM is able to outperform  the SPM. 

In Case study 2, the CPM clearly outperformed the SPM. The Lotka-Volterra model, System \eqref{__label__f677e98bcf0242e29d719f07a0be9356}, was a particularly challenging test case, and we follow \citet{__ref__77854e0e556e40b2bec0a7e868cb9542} in explaining why. \citet{__ref__77854e0e556e40b2bec0a7e868cb9542} argues that if an ODE modelling approach is followed, then, in the $X_1$-$X_2$ plane, the solution trajectories of System \eqref{__label__f677e98bcf0242e29d719f07a0be9356} are closed orbits. \citet{__ref__77854e0e556e40b2bec0a7e868cb9542} then says that the addition of microscopic fluctuations (due to using a stochastic model) induces a `drunkard's walk' over the continuum of deterministic orbits, thereby resulting in unstable behaviour. This feature makes it difficult to ensure that pairs of sample paths are tightly coupled. In particular, if a pair of sample paths differs slightly in their state vectors, then the difference in state vectors continues to increase with the SPM method. The CPM does not suffer the same defect, and, as such, it is more efficient than the DM. 

In Case study 3, a case study of a logistic growth model was presented, and the performance of the multi-level method was evaluated. {The relative performance of the CPM and SPM depended on the choice of estimator: $\mathcal{Q} = \mathbb{E}[X(T)]$ or $\mathcal{Q} = \mathbb{E}[\int_0^T X(t)\wrt{t}/T]$. The SPM is more efficient for the former statistic, and the CPM for the latter. We posit that, by time $T=3$, the transient dynamics of the system are no longer relevant: $X$ has increased rapidly in value, and is now fluctuating rapidly about a steady state. We suggest that the ``memoryless'' property of the SPM means it is able to estimate $\mathbb{E}[X(T)]$ very efficiently. Where $\mathbb{E}[\int_0^TX(t)\wrt{T} /T]$ is estimated, stochastic variations in the transient dynamics have a greater effect on the quantity of interest, and the ``path-dependent'' CPM method is very efficient. 

Finally, in Case study 4 we demonstrated the performance of the multi-level method with a relatively complicated stochastic model. We showed that, for this system, both the CPM and SPM implementations outperform regular DM simulation, with the CPM multi-level method being nearly 15\% faster than the SPM technique. }

\subsection{Comparing tau-leap and R-leap multi-level methods}
By treating the time traversed by each algorithm step of the R-leap method, and the particular combination of reactions that take place during that step, as two distinct and unrelated quantities that must be determined by Monte Carlo simulation, we have described a new variance reduction technique for multi-level simulation. The R-leap multi-level method was assessed by considering Case study 3. We demonstrated that the R-leap approach performed slightly better than the CPM-driven, tau-leap multi-level method, but that it did not perform as well as the SPM-driven method. In summary, the R-leap multi-level method has the potential to provide good computational performance.

The crucial difference between the tau-leap and R-leap multi-level methods is seen by comparing Equations \eqref{__label__8b0e5f6d7021402b8ab26991f440427d} and \eqref{__label__458ab95eeeac4b7dbba9920ef4b5682f}. The tau-leap method couples two distinct sample paths by considering the difference in \emph{absolute propensity values} of corresponding reaction channels in the sample paths, whilst the R-leap method considers the difference in the corresponding \emph{proportion of the total propensity} (or probability) attributable to each reaction channel. Variance reduction in the tau-leap and R-leap multi-level techniques arises in a different format, and the computational performance is therefore different.

\subsection{Outlook}
The CPM provides a natural framework for implementing the multi-level method. It is able to mitigate some of the difficulties associated with the previously-used SPM implementation. Future work will include categorising reaction networks and summary statistics in order to derive criteria to decide whether the CPM or SPM should be used for that particular problem. Hybrid approaches that combine the SPM and CPM can also be implemented~\citep{__ref__ed306348b1fd4394bdcb1fd9647a6f77}. The R-leap method has been successfully implemented, and future work will determine the problems it is most suited to handling. Ultimately, a refined multi-level method will dramatically reduce the computational burden of Monte Carlo simulation. 

\newpage
\section*{Bibliography}

\bibliographystyle{v2}
\bibliography{refs_clean.bib}

\begin{thebibliography}{25}
\providecommand{\natexlab}[1]{#1}

\bibitem[Hanna et~al.(2009)]{__ref__c9c0acbfcc1e466796f203c599f87a48}
Hanna, J., Saha, K., Pando, B., Van~Zon, J., Lengner, C.~J., Creyghton, M.~P.,
  van Oudenaarden, A., and Jaenisch, R.
\newblock Direct cell reprogramming is a stochastic process amenable to
  acceleration.
\newblock \emph{Nature}, \textbf{462}(7273):595--601, 2009.

\bibitem[Lande et~al.(2003)]{__ref__8a7bc56c9dfd469dbc93e2e8bbf7172b}
Lande, R., Engen, S., and Saether, B.-E.
\newblock \emph{Stochastic Population Dynamics in Ecology and Conservation}.
\newblock Oxford University Press, 2003.

\bibitem[Thattai and
  Van~Oudenaarden(2001)]{__ref__fa6e3de85e0b4412ba3096011a9acb70}
Thattai, M. and Van~Oudenaarden, A.
\newblock Intrinsic noise in gene regulatory networks.
\newblock \emph{Proceedings of the National Academy of Sciences},
  \textbf{98}(15):8614--8619, 2001.

\bibitem[Van~Kampen(1992)]{__ref__37a69023cd02435cb5c677546fc5d00e}
Van~Kampen, N.~G.
\newblock \emph{Stochastic Processes in Physics and Chemistry}.
\newblock Elsevier, 1992.

\bibitem[Gillespie et~al.(2013)]{__ref__3e0b45aec9df4e05b2f3e8165dab4878}
Gillespie, D.~T., Hellander, A., and Petzold, L.~R.
\newblock Perspective: Stochastic algorithms for chemical kinetics.
\newblock \emph{Journal of Chemical Physics}, \textbf{138}(17):170901, 2013.

\bibitem[Giles(2008)]{__ref__dbd754638429454ab66f23e25ec8f0be}
Giles, M.~B.
\newblock {Multilevel Monte Carlo path simulation}.
\newblock \emph{Operations Research}, \textbf{56}(3):607--617, 2008.

\bibitem[Anderson and Higham(2012)]{__ref__5f82fa75a5314b0484ebeb52de9e2750}
Anderson, D. and Higham, D.
\newblock Multi-level Monte Carlo for continuous time Markov chains, with
  applications in biochemical kinetics.
\newblock \emph{SIAM Multiscale Modeling and Simulation},
  \textbf{10}(1):146--179, 2012.

\bibitem[Lester et~al.(2015)]{__ref__cefabb61e12c4615ba29a64649c34236}
Lester, C., Yates, C.~A., Giles, M.~B., and Baker, R.~E.
\newblock An adaptive multi-level simulation algorithm for stochastic
  biological systems.
\newblock \emph{Journal of Chemical Physics}, \textbf{142}(2):024113, 2015.

\bibitem[Moraes et~al.(2016)]{__ref__25cacfb38a074fdf9ed2a1329b16f350}
Moraes, A., Tempone, R., and Vilanova, P.
\newblock Multilevel hybrid Chernoff tau-leap.
\newblock \emph{BIT Numerical Mathematics}, \textbf{56}(1):189--239, 2016.

\bibitem[Hammouda et~al.(2017)]{__ref__32e9156a6baf4b31a6cfc19f7b417e76}
Hammouda, C.~B., Moraes, A., and Tempone, R.
\newblock Multilevel hybrid split-step implicit tau-leap.
\newblock \emph{Numerical Algorithms}, \textbf{74}(2):527--560, 2017.

\bibitem[Lester et~al.(2016)]{__ref__b7cd69d78ffc4c2ba54f2f3e26b5d981}
Lester, C., Baker, R.~E., Giles, M.~B., and Yates, C.~A.
\newblock Extending the multi-level method for the simulation of stochastic
  biological systems.
\newblock \emph{Bulletin of Mathematical Biology}, \textbf{78}(8):1640–1677,
  2016.

\bibitem[Kurtz(1980)]{__ref__93139130af66492c89b2da93df8e8526}
Kurtz, T.~G.
\newblock Representations of Markov processes as multiparameter time changes.
\newblock \emph{Annals of Probability}, pages 682--715, 1980.

\bibitem[Schnoerr et~al.(2017)]{__ref__31cfa34968c84ce9be58a56caac7157e}
Schnoerr, D., Sanguinetti, G., and Grima, R.
\newblock Approximation and inference methods for stochastic biochemical
  kinetics—a tutorial review.
\newblock \emph{Journal of Physics A: Mathematical and Theoretical},
  \textbf{50}(9):093001, 2017.

\bibitem[Gillespie(1977{\natexlab{a}})]{__ref__85bbd2b5b7ad4596ad0a5acfb3b44152}
Gillespie, D.~T.
\newblock {Exact stochastic simulation of coupled chemical reactions}.
\newblock \emph{Journal of Physical Chemistry}, \textbf{81}(25):2340--2361,
  1977{\natexlab{a}}.

\bibitem[Gillespie(1976)]{__ref__9e06f02ea6d8426bb0cae90cddbbe6e7}
Gillespie, D.~T.
\newblock {A general method for numerically simulating the stochastic time
  evolution of coupled chemical reactions}.
\newblock \emph{Journal of Computational Physics}, \textbf{22}(4):403--434,
  1976.

\bibitem[Gillespie(2001)]{__ref__c646c0890ccf4a59835f2df5c824c223}
Gillespie, D.
\newblock Approximate accelerated stochastic simulation of chemically reacting
  systems.
\newblock \emph{{Journal of Chemical Physics}}, \textbf{115}(4):1716--1733,
  2001.

\bibitem[Auger et~al.(2006)]{__ref__d45dab7799794c30abe917c6253324aa}
Auger, A., Chatelain, P., and Koumoutsakos, P.
\newblock {R-leaping: accelerating the stochastic simulation algorithm by
  reaction leaps.}
\newblock \emph{Journal of Chemical Physics}, \textbf{125}(8):084103, 2006.

\bibitem[Li(2007)]{__ref__d5682854157f4e11b42663588059cb23}
Li, T.
\newblock {Analysis of explicit tau-leaping schemes for simulating chemically
  reacting systems}.
\newblock \emph{SIAM Multiscale Modeling and Simulation},
  \textbf{6}(2):417--436, 2007.

\bibitem[Rathinam et~al.(2010)]{__ref__06e05ee3d1954ff58fd9cb7791b406de}
Rathinam, M., Sheppard, P.~W., and Khammash, M.
\newblock {Efficient computation of parameter sensitivities of discrete
  stochastic chemical reaction networks.}
\newblock \emph{Journal of Chemical Physics}, \textbf{132}(3):034103, 2010.

\bibitem[Anderson(2007)]{__ref__54e4fec61bb749dfacb9f7a516266af0}
Anderson, D.
\newblock A modified next reaction method for simulating chemical systems with
  time-dependent propensities and delays.
\newblock \emph{Journal of Chemical Physics}, \textbf{127}(21):214107, 2007.

\bibitem[Murray(2002)]{__ref__39b5170e8a0949e99c877921832cdd3f}
Murray, J.
\newblock \emph{Mathematical Biology}, volume~2.
\newblock Springer, 2002.

\bibitem[Gillespie(1977{\natexlab{b}})]{__ref__77854e0e556e40b2bec0a7e868cb9542}
Gillespie, D.
\newblock Exact stochastic simulation of coupled chemical reactions.
\newblock \emph{{Journal of Physical Chemistry}}, \textbf{81}(25):2340--2361,
  1977{\natexlab{b}}.

\bibitem[MacNamara et~al.(2008)]{__ref__6c309c4c12884051a38b7be20b047e02}
MacNamara, S., Bersani, A.~M., Burrage, K., and SiDesmond~Je, R.~B.
\newblock Stochastic chemical kinetics and the total quasi-steady-state
  assumption: application to the stochastic simulation algorithm and chemical
  master equation.
\newblock \emph{Journal of Chemical Physics}, \textbf{129}(9):095105, 2008.

\bibitem[Huang and Ferrell(1996)]{__ref__cfcfd0d4f21146a3bb4809687529f7702}
Huang, C.-Y. and Ferrell, J.~E.
\newblock Ultrasensitivity in the mitogen-activated protein kinase cascade.
\newblock \emph{Proceedings of the National Academy of Sciences},
  \textbf{93}(19):10078--10083, 1996.

\bibitem[Lester et~al.(2017)]{__ref__ed306348b1fd4394bdcb1fd9647a6f77}
Lester, C., Yates, C.~A., and Baker, R.~E.
\newblock Efficient parameter sensitivity computation for spatially-extended
  reaction networks.
\newblock \emph{Journal of Chemical Physics}, \textbf{146}(4):044106, 2017.

\end{thebibliography}

\end{document}